\documentclass[aps, prc,
reprint,
longbibliography, 
nofootinbib,
superscriptaddress,
preprintnumbers
]{revtex4-2}

\makeatletter
\def\@bibdataout@aps{%
\immediate\write\@bibdataout{%
@CONTROL{%
apsrev41Control%
\longbibliography@sw{%
    ,author="08",editor="1",pages="1",title="0",year="1"%
    }{%
    ,author="08",editor="1",pages="1",title="",year="1"%
    }%
  }%
}%
\if@filesw \immediate \write \@auxout {\string \citation {apsrev41Control}}\fi
}
\makeatother

\usepackage{setspace}

\usepackage{amsfonts}
\usepackage{amsmath}
\usepackage{amssymb}

\DeclareFontFamily{OT1}{pzc}{}
\DeclareFontShape{OT1}{pzc}{m}{it}{<-> s * [1.10] pzcmi7t}{}
\DeclareMathAlphabet{\mathpzc}{OT1}{pzc}{m}{it}

\usepackage{booktabs}
\AtBeginDocument{
\heavyrulewidth=.08em
\lightrulewidth=.05em
\cmidrulewidth=.03em
\belowrulesep=.65ex
\belowbottomsep=0pt
\aboverulesep=.4ex
\abovetopsep=0pt
\cmidrulesep=\doublerulesep
\cmidrulekern=.5em
\defaultaddspace=.5em
}

\newcommand{\doubletoprule}[1]%
  {\toprule\multicolumn{#1}{c}{}\\[-1.3em]\toprule}
\newcommand{\doublebottomrule}[1]%
  {\bottomrule\multicolumn{#1}{c}{}\\[-1.2em]\bottomrule}

\usepackage{pgfplots, pgfplotstable}

\usepackage{cancel}
\usepackage{caption}
\usepackage{chemformula} 
\usepackage{comment} 
\usepackage[inline]{enumitem}
\usepackage[nowatermark]{fixmetodonotes}
\usepackage{float}
\usepackage{graphicx}
\usepackage{isotope}
\usepackage{makecell} 
\usepackage{mathtools} 
\usepackage{multirow} 
\usepackage{siunitx}
\usepackage{stackrel} 
\DeclareSIUnit\ton{t}
\DeclareSIUnit\parsec{pc}
\DeclareSIUnit[number-unit-product = ]\percent{\char`\%}

\usepackage{threeparttable} 
\usepackage{xifthen} 

\usepackage[colorlinks, urlcolor=blue, citecolor=blue, menucolor=.,
  anchorcolor=., linkcolor=., runcolor=.]{hyperref}

\DeclareMathOperator{\Tr}{Tr}
\DeclareMathOperator{\sgn}{sgn}

\usepackage{orcidlink}

\newcommand{\chirality}{\xi}

\newcommand{\pnu}{k}
\newcommand{\plep}{\pnu^\prime}
\newcommand{\pNi}{p}
\newcommand{\pNf}{\pNi^\prime}

\newcommand{\Lep}{\mathsf{L}}

\newcommand{\RLep}{\mathsf{v}}

\newcommand{\Nuc}{\mathsf{W}}

\newcommand{\RNuc}{\mathsf{R}}

\newcommand{\LepFunc}{v}

\newcommand{\NucResp}{R}

\newcommand{\NucMat}{\mathcal{N}}

\newcommand{\IndexForWalecka}{\mathcal{X}}
\newcommand{\WaleckaMatrixEl}{W}

\newcommand{\nucIndex}{n}

\newcommand{\CoulombFactor}{F_C}

\newcommand{\SquaredAmplitude}{|\overline{\mathcal{M}}|^2}

\newcommand{\FNRpLep}{\mathpzc{K}}
\newcommand{\FNReLep}{\mathpzc{E}}
\newcommand{\FNRpLepEff}{\FNRpLep_\mathrm{\,\,eff}}
\newcommand{\FNReLepEff}{\FNReLep_\mathrm{\,eff}}

\newcommand{\FermiMat}{B(\mathrm{F})}

\newcommand{\GTMat}{B(\mathrm{GT})}

\newcommand{\LorT}{t}
\newcommand{\LorX}{x}
\newcommand{\LorY}{y}
\newcommand{\LorZ}{z}

\renewcommand{\Re}{\operatorname{Re}}
\renewcommand{\Im}{\operatorname{Im}}

\newcommand{\BinIdx}{\ensuremath{i}}

\newcommand{\FSIdx}{\ensuremath{F}}

\newcommand{\BR}{\ensuremath{\mathcal{R}}}

\newcommand{\DeltaCRPA}{\ensuremath{\Delta_\mathrm{CC}}}

\newcommand{\Eff}{\ensuremath{\mathrm{eff}}}

\newcommand{\omegaEff}{\ensuremath{\omega_\Eff}}

\newcommand{\RNucWide}{\widetilde{\RNuc}}

\newcommand{\RNucEff}{\RNucWide_\Eff}

\newcommand{\ExContinuumThresh}{\ensuremath{E_x^\mathrm{c}}}

\newcommand{\NucRespWide}{\widetilde{\NucResp}}

\newcommand{\omegaWide}{\omega^\prime}

\newcommand{\LorentzianWidth}{\Gamma_\mathrm{eff}}

\newcommand{\LorentzianNorm}{C_L}

\newcommand{\FragIdx}{\Lambda}

\newcommand{\mAtom}{m_\mathrm{atom}}

\newcommand{\mNuclear}{m_\mathrm{nucl}}

\newcommand{\quadA}{\mathpzc{A}}
\newcommand{\quadB}{\mathpzc{B}}
\newcommand{\quadC}{\mathpzc{C}}
\newcommand{\quadD}{\mathpzc{D}}

\begin{document}

\preprint{FERMILAB-PUB-26-0249-CSAID-STUDENT-T}


\newcommand{\marley}{\texttt{MARLEY}}

\newcommand{\version}{2.0.0}

\newcommand{\oldVersion}{1.2.0}

\newcommand{\unaryminus}{\scalebox{0.5}[0.72]{\( - \!\)}}

\newcommand{\marleyOld}{\texttt{v1}}
\newcommand{\marleyNew}{\texttt{v2}}

\title{Continuum contribution to charged-current absorption of low-energy
$\nu_e$ on $^{40}$Ar}

\author{Steven Gardiner\,\orcidlink{0000-0002-8368-5898}\,}
\email{gardiner@fnal.gov}
\affiliation{Fermi National Accelerator Laboratory, Batavia,
  Illinois 60510, USA}

\author{Pablo Barham Alz\'{a}s\,\orcidlink{0000-0001-9640-8219}\,}
\thanks{Present address: Tel Aviv University, Tel Aviv 69978, Israel}
\affiliation{CERN, The European Organization for
  Nuclear Research, 1211 Meyrin, Switzerland}

\author{Alexis Nikolakopoulos\,\orcidlink{0000-0001-6963-8115}\,}
\affiliation{Department of Physics, University of Washington, Seattle, Washington 98195, USA}
\affiliation{Department of Physics and Astronomy, Ghent University,
  Proeftuinstraat 86, B-9000 Gent, Belgium}

\author{Luca H. Abu El-Haj\,\orcidlink{0009-0005-5021-4212}\,}
\affiliation{Fermi National Accelerator Laboratory, Batavia,
Illinois 60510, USA}
\affiliation{Columbia University, New York, New York 10027, USA}

\author{Natalie Jachowicz\,\orcidlink{0000-0003-1168-0745}\,}
\affiliation{Department of Physics and Astronomy, Ghent University,
  Proeftuinstraat 86, B-9000 Gent, Belgium}

\author{Vishvas Pandey\,\orcidlink{0000-0002-3082-7987}\,}
\affiliation{Fermi National Accelerator Laboratory, Batavia,
  Illinois 60510, USA}

\date{\today}

\begin{abstract}
\begin{description}
\item[Background] Accurate modeling of the absorption of tens-of-MeV $\nu_e$ on
$^{40}$Ar is needed to enable measurements of astrophysical neutrinos using
large liquid argon time projection chamber (LArTPC) detectors, such as those
planned for the Deep Underground Neutrino Experiment (DUNE).
\item[Purpose] We revisit the \marley\ neutrino interaction model used in
present estimates of DUNE sensitivity to supernova and solar neutrino signals.
Multiple theoretical refinements are pursued, especially in the unbound
continuum region of nuclear excitation energy.
\item[Methods] Inclusive charged-current neutrino-argon cross sections are
calculated using a hybrid strategy. Nuclear transitions to unbound states are
treated using a Hartree-Fock Continuum Random Phase Approximation (HF-CRPA)
model, including forbidden contributions. Allowed transitions to low-lying
discrete levels are also included using indirect measurements and approximate
corrections for the momentum transfer dependence. Exclusive predictions are
obtained by coupling these calculations with a statistical nuclear
de-excitation model.
\item[Results] The impact on observables of interest for DUNE and similar
experiments is examined in terms of both total and differential cross sections.
Our refined calculations predict a lower allowed portion of the cross section
relative to the prior \marley\ model. At neutrino energies appreciably below
100~MeV, the inclusion of forbidden transitions does not fully compensate for
the loss of allowed strength.
\item[Conclusions] For a representative neutrino burst from a galactic
core-collapse supernova, our results suggest that \marley~\oldVersion\
overestimates the event yield in a DUNE-like detector by
approximately~20\%. However, because this overestimation is more severe at
backwards angles, use of the charged-current $\nu_e$-$^{40}$Ar reaction for
supernova pointing may be more feasible than previously expected.
\end{description}
\end{abstract}


\maketitle

\section{Introduction}

Ever since the first detection of two dozen supernova neutrinos in
1987~\cite{Hirata1987, Alekseev1987, Bionta1987}, there has been sustained
exploration of their implications and of the scientific potential of future,
more detailed observations~\cite{Raffelt1990, Schaeffer1990, Jegerlehner1996,
Vissani2014, Mirizzi2016, Branch2017}. Efforts are ongoing to maximize
experimental sensitivity to the next galactic core-collapse
supernova~\cite{Antonioli2004, Scholberg2008, AlKharusi2021} as well as to
detect the \textit{diffuse supernova neutrino background} (DSNB), a small
continuous flux of neutrinos generated by all past explosions~\cite{Beacom2010,
marleyTheoryExample1}. The physics program that would be enabled by
next-generation measurements of supernova neutrinos is wide-ranging and
discussed at length in multiple recent reviews~\cite{Muller2019, Horiuchi2018,
Raffelt2011}. Because the interior of a supernova is opaque to photons
(delaying their ultimate emission by hours), weakly-interacting neutrinos
provide a unique probe of early stages of stellar collapse. A supernova
observation could also be used to determine the neutrino mass
ordering~\cite{Scholberg2018} and search for evidence of a variety of
hypothetical non-standard interactions~\cite{Farzan2003, Pretel2007,
Stapleford2016, Akita2022}. A topic of intense recent
attention~\cite{Patwardhan2020, Akhmedov2017, Mirizzi2016, Duan2010} that mixes
both astrophysics and particle physics is \textit{collective neutrino
oscillations}: in the extreme conditions created by a core-collapse supernova,
neutrino-neutrino scattering becomes an important process, leading to
complicated behavior potentially detectable as features in the time-dependent
energy spectra of the emitted neutrinos.

Maximizing the scientific impact of next-generation supernova observations will
require measuring neutrinos of all flavors~\cite{Nikrant2018, Boyd2003}. Thanks
to the neutron excess in \isotope[40]{Ar} as well as their detailed tracking
capabilities, liquid argon time projection chambers (LArTPCs) are expected to be
uniquely capable of measuring supernova $\nu_e$ with high statistics and minimal
backgrounds~\cite{Abi2020, Ankowski2016, Scholberg2012}. The ten-kiloton-scale
LArTPCs that will begin operation later this decade as the far detector of the
Deep Underground Neutrino Experiment (DUNE) will thus be an invaluable addition
to a worldwide network of supernova neutrino observatories~\cite{Abi2020,
duneTDRvol4}. The DUNE LArTPCs may also enable unique measurements of solar
neutrinos~\cite{DUNEsolar, Meighen-Berger:2024xbx} and competitive searches for
cosmic neutrinos generated by exotic mechanisms, such as dark matter
decay~\cite{DeRomeri:2021xgy, Buckley:2022btu}.

At the tens-of-MeV energies relevant for supernova neutrinos, many nuclear
structure details must be considered to provide a realistic description of
inelastic neutrino scattering on complex nuclei. Inclusive neutrino-nucleus
cross sections in this regime receive significant contributions from
transitions to individual low-lying nuclear energy levels as well as from giant
resonances which predominantly populate the unbound continuum. A complete
interaction model must couple predictions of these contributions with a
treatment of secondary particle emission from the outgoing nucleus. Because
DUNE's ability to reconstruct supernova neutrino energies will be sensitive to
how energy is partitioned between final-state particle
species~\cite{marleyPRC}, such a complete interaction model will be essential
to interpret a future observation.

Faced with a similar need for precise neutrino energy reconstruction despite
nuclear theory challenges of their own~\cite{alvarezruso2018}, experimental
collaborations working with GeV accelerator neutrino beams rely upon
simulations of complete neutrino-nucleus interactions provided by Monte Carlo
event generators such as Achilles~\cite{achilles2023, Isaacson:2021xty},
GENIE~\cite{genie2021, andreopoulos2010}, GiBUU~\cite{Mosel:2023zek,
Buss:2011mx}, NEUT~\cite{hayato2009, hayato2021}, and NuWro~\cite{golan2012}.
In light of significant differences between the treatments of nuclear physics
appropriate for the MeV and GeV scales, a dedicated event generator for
low-energy neutrinos called \marley\ (Model of Argon Reaction Low Energy
Yields) was recently created~\cite{marleyCPC}.

Version~\oldVersion\ of \marley\footnote{The newer 1.2.1 release, which is the
latest public version as of this writing, includes a small technical fix that
allows \marley\ to run concurrently with \texttt{BxDecay0}~\cite{bxdecay0} in
the DUNE simulation chain. The physics model is unchanged from
\marley~\oldVersion.} implements a neutrino scattering model~\cite{marleyPRC}
for the charged-current (CC) reaction
\begin{equation}
\nu_e + \isotope[40]{Ar} \to e^- + \isotope[40]{K}^*
\end{equation}
and subsequent nuclear de-excitations. This model currently serves as the
foundation for official estimates of DUNE sensitivity to supernova
neutrinos~\cite{DUNE:2020zfm, xsecuncdunesn, DUNE:2024ptd}, with studies of the
DSNB and solar neutrinos also underway~\cite{CuestaSoria:2023gss,
MantheyCorchado:2024nao}. Beyond DUNE, the code has been directly used to model
both signal~\cite{ninspb, coherent:2023ffx} and background~\cite{csicoherent}
in analyses by the COHERENT experiment, to study supernova $\nu_e$ sensitivity
in the nEXO experiment~\cite{nEXO:2024uyw}, and to carry out a variety of
simulation-based studies of theoretical topics~\cite{Buckley:2022btu,
Martinez-Soler:2021unz, DeGouvea:2020ang, deGouvea:2019goq} as well as novel
detector concepts and experimental techniques~\cite{castiglioni2020, qpix,
Hartsock:2025ipn, Folkerts:2023svx, Mastbaum:2022rhw, Shi:2025rob}.
Cross-section predictions originally produced with \marley\ have also seen
frequent use as a reference model for many phenomenological investigations
related to astrophysical neutrinos~\cite{DeRomeri:2021xgy,
Meighen-Berger:2024xbx, Ge:2024ftz, Saez:2024ayk, Saez:2023snv, pompa2022,
kelly2021, Bendahman:2021zxz, Li:2020ujl}.

While \marley's distinctive approach to MeV-scale interaction simulations has
motivated its widespread adoption for these tasks, the neutrino cross-section
model provided in version~\oldVersion\ is based on a calculation strategy called
the \textit{allowed approximation} (AA) with known and significant theoretical
shortcomings~\cite{VanDessel:2019atx, VanDessel:2019obk}. To overcome these
shortcomings and provide a higher-quality simulation for DUNE and other
interested parties, we have removed the AA from \marley\ using complementary
approaches to improve the theoretical treatment for discrete and continuum
nuclear transitions.

In this article, we present the refined cross-section calculations developed
for the upcoming public release of version~\version\ of \marley.
Section~\ref{sec:model} discusses the physics model improvements and explains
the limitations of the previous AA-based approach. Section~\ref{sec:results}
reports many calculated cross sections obtained with the new \marley~\version\
simulation, highlighting some noteworthy changes from previous predictions.
Finally, Sec.~\ref{sec:summary} provides a brief summary and conclusions.

\section{Model}
\label{sec:model}

For momentum transfers that are small compared to the $W$ boson mass, the
tree-level amplitude $\mathcal{M}$ for inclusive charged-current
neutrino-nucleus scattering may be represented diagrammatically as
\begin{equation}
i\mathcal{M} = 
\vcenter{\hbox{ \includegraphics{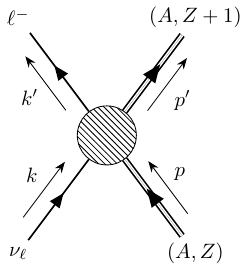} }} \,.
\end{equation}
The squared amplitude, averaged over the initial nuclear spin and summed over
final spins, may be written in the form
\begin{equation}
\label{eq:CC_squared_amplitude}
\SquaredAmplitude = \frac{ G_F^2 \, |V_{ud}|^2 }{ 2 }
  \, \CoulombFactor \, \Lep_{\mu\nu} \, \Nuc^{\mu\nu}
\end{equation}
where $G_F$ is the Fermi constant, $V_{ud}$ is the Cabibbo–Kobayashi–Maskawa
matrix element connecting the up and down quarks, and $\CoulombFactor$ is the
Coulomb correction factor discussed in Sec.~\ref{sec:coulomb_corrections}
below. The leptonic ($\Lep_{\mu\nu}$) and hadronic ($\Nuc^{\mu\nu}$) tensors
are defined by
\begin{align}
\label{eq:CC_lepton_tensor}
\Lep_{\mu\nu} &\equiv \Tr[\gamma_\mu \, (1 - \gamma_5) \, \cancel{\pnu} \,
\gamma_{\nu} \, (1 - \gamma_5) \, (\cancel{\pnu}^\prime + m_\ell)]
\\ & \!\!\!\!\!= 8\left[ \pnu_\mu \, \plep_\nu
+ \pnu_\nu \, \plep_\mu
- g_{\mu \nu}\, (\pnu \cdot \plep)
+ i \, \chirality \, \epsilon_{\mu \nu \rho \sigma} \, {\plep}^\rho \pnu^\sigma
\right]
\end{align}
and
\begin{equation}
\label{eq:CC_nuclear_tensor}
\Nuc^{\mu\nu} \equiv \frac{1}{2J_i + 1} \sum_{M_i} \sum_{M_f}
\NucMat^\mu\,\NucMat^{\nu*} \,,
\end{equation}
where $m_\ell$ is the mass of the final-state lepton, $\epsilon_{\mu \nu \rho
\sigma}$ is the Levi-Civita symbol, $g_{\mu \nu}$ is the metric tensor, $\pnu$
is the four-momentum of the incoming (anti)neutrino, and $\plep$ is the
four-momentum of the outgoing lepton. The chirality of the incident particle is
$\chirality = -1$ for neutrinos and $\chirality = +1$ for antineutrinos. The
symbol $J_i$ ($J_f$) represents the initial (final) nuclear spin, and $M_i$
($M_f$) denotes the third component of the nuclear spin in the initial (final)
state.

Under the impulse approximation, the nuclear matrix element may be written in
coordinate space as
\begin{equation}
\label{eq:nuclear_matrix_element}
\NucMat^\mu =
\sqrt{2E_i}\,\sqrt{2E_f}\,
\big< J_f \big| \, {\textstyle\sum_{\nucIndex=1}^A} \, e^{ i \mathbf{q}
\cdot \mathbf{r}_\nucIndex } \, j^\mu(\nucIndex) \, \big| J_i \big>
\end{equation}
where $E_i$ ($E_f$) is the total energy of the nucleus in the initial (final)
state, $\mathbf{q} = \mathbf{\pnu} - \mathbf{\plep} = \mathbf{\pNf} -
\mathbf{\pNi}$ is the three-momentum transfer, and the sum runs over all $A$
nucleons. The weak current operator $j^\mu(\nucIndex)$ is understood to act
only on the $\nucIndex$th nucleon, which has the position vector
$\mathbf{r}_\nucIndex$. The state vectors labeled with the nuclear spin are
normalized to unity, i.e.,
\begin{equation}
\big< J_i \big| J_i \big> =
\big< J_f \big| J_f \big> = 1 \,.
\end{equation}

\subsection{Tensor contraction}
\label{sec:tensor_contraction}

The tensor contraction that appears in Eq.~\ref{eq:CC_squared_amplitude} may be
rewritten in the form
\begin{equation}
\nonumber \Lep_{\mu \nu} \, \Nuc^{\mu \nu} = 32 \, E_\nu \, E_\ell \, E_i \, E_f
\cdot \RLep_{\mu \nu} \, \RNuc^{\mu \nu} \,,
\end{equation}
where the dimensionless leptonic ($\RLep_{\mu \nu}$) and hadronic ($\RNuc^{\mu
\nu}$) tensors are defined by
\begin{align}
\RLep_{\mu \nu} &\equiv \frac{ \Lep_{\mu \nu} }{8 \, E_\nu \, E_\ell } &
\RNuc^{\mu \nu} &\equiv \frac{ \Nuc^{\mu \nu} }{ 4 \, E_i \, E_f } \,,
\end{align}
and $E_\nu$, $E_\ell$, $E_i$, and $E_f$ are the total energies of the
incoming neutrino, outgoing lepton, initial nucleus, and final nucleus,
respectively. The new tensor contraction may, in turn, be written as
\begin{align}
\nonumber \RLep_{\mu \nu} \, \RNuc^{\mu \nu}
= \LepFunc_{CC} &\, \NucResp_{CC} + \LepFunc_{CL} \, \NucResp_{CL}
+ \LepFunc_{LL} \, \NucResp_{LL}
\\ \label{eq:new_contraction} &+ \LepFunc_{T} \, \NucResp_{T}
+ \LepFunc_{T^\prime} \, \NucResp_{T^\prime} \,,
\end{align}
To discuss the variables on the right-hand side of
Eq.~\ref{eq:new_contraction} generically, we introduce the
index
\begin{equation}
  \IndexForWalecka \in \{ CC, CL, LL, T, T^\prime \} \,.
\end{equation}
The $\LepFunc_\IndexForWalecka$ factors are related to
the elements of $\RLep_{\mu \nu}$ by the expressions
\begin{align}
\label{eq:first_v} \LepFunc_{CC} &= \RLep_{\LorT \LorT} \\
\LepFunc_{CL} = -\Re(\RLep_{\LorT \LorZ})
  &= -\frac{1}{2}\big(\RLep_{\LorT \LorZ}
+ \RLep_{\LorZ \LorT} \big) \\
\LepFunc_{LL} &= \RLep_{\LorZ \LorZ} \\
\LepFunc_{T} &= \RLep_{\LorX \LorX} + \RLep_{\LorY \LorY} \\
\label{eq:last_v} \LepFunc_{T^\prime}
  = \Im(\RLep_{\LorX \LorY})
  &= -\frac{i}{2}
\big( \RLep_{\LorX \LorY} - \RLep_{\LorY \LorX} \big)
\end{align}
where we have noted that $\RLep_{\mu \nu}$ is Hermitian ($\RLep_{\mu \nu} =
\RLep_{\nu \mu}^{*}$). The expressions in
Eqs.~\ref{eq:first_v}--\ref{eq:last_v} also hold for the nuclear response
functions $\NucResp_\IndexForWalecka$ written in terms of elements of the
dimensionless hadronic tensor, e.g.,
\begin{equation}
\NucResp_{CC} = \RNuc^{\LorT \LorT} \,.
\end{equation}

In a frame where $\mathbf{q}$ points along the positive $z$ direction,
the $v_\IndexForWalecka$ factors become
\begin{align}
\label{eq:first_lep_func}
\LepFunc_{CC} &= 1 + \beta_\ell \, \cos\theta_\ell \\
\LepFunc_{CL} &= -\bigg[ \frac{ \omega }{ \kappa }
  \big(1 + \beta_\ell \, \cos\theta_\ell \big)
  + \frac{ m_\ell^2 }{ E_\ell \, \kappa } \bigg] \\
\LepFunc_{LL} &= 1 + \beta_\ell \cos\theta_\ell
- \frac{ 2\,E_\nu\,E_\ell }{ \kappa^2 } \beta_\ell^2 \sin^2\theta_\ell \\
\LepFunc_{T} &= 1 - \beta_\ell \cos\theta_\ell
+ \frac{ E_\nu\,E_\ell }{ \kappa^2 } \beta_\ell^2 \sin^2\theta_\ell \\
\label{eq:last_lep_func}
\LepFunc_{T^\prime} &= \chirality \, \bigg[ \frac{ E_\nu + E_\ell }{ \kappa }
  \big( 1 - \beta_\ell \cos\theta_\ell \big)
  - \frac{ m_\ell^2 }{ E_\ell \, \kappa } \bigg]
\end{align}
where $\omega = E_\nu - E_\ell$ is the energy transfer, $\kappa = |\mathbf{q}|$
is the magnitude of the three-momentum transfer, $\beta_\ell =
|\mathbf{\plep}| / E_\ell$ is the speed of the outgoing lepton, and
$\theta_\ell$ is the scattering angle, i.e.,
\begin{equation}
\theta_\ell = \arccos\bigg( \frac{ \mathbf{\pnu} \cdot \mathbf{\plep} }
{ |\mathbf{\pnu}| \, |\mathbf{\plep}| } \bigg) \,.
\end{equation}

\subsection{Nuclear responses}
\label{sec:nuclear_responses}

For the same choice of frame, the nuclear responses may be written in the form
\begin{equation}
\label{eq:NucResponseFunc}
\NucResp_\IndexForWalecka = \frac{ 4\pi }{ 2J_i + 1 }
  \sum_{J=J_\mathrm{min}^\IndexForWalecka}^\infty
\WaleckaMatrixEl_\IndexForWalecka
\end{equation}
where $J$ is the multipole order,
\begin{equation}
J_\mathrm{min}^\IndexForWalecka \equiv \begin{cases}
0 & \IndexForWalecka \in \{ CC, CL, LL \} \\
1 & \IndexForWalecka \in \{ T, T^\prime \}
\end{cases} \,,
\end{equation}
and the $\WaleckaMatrixEl_\IndexForWalecka$
are the Walecka~\cite{Walecka1975} reduced matrix elements:
\begin{align}
\label{eq:first_Walecka_ME}
\WaleckaMatrixEl_{CC} &= \big| \big< J_f \big\Vert
  \mathcal{\hat{M}}_J \big\Vert J_i \big> \big|^2
\\
\WaleckaMatrixEl_{CL} &= -2 \Re\!\Big[ \big< J_f \big\Vert
  \mathcal{\hat{M}}_J \big\Vert J_i \big>
\big< J_f \big\Vert
  \mathcal{\hat{L}}_J \big\Vert J_i \big>^{*}
\Big]
\\
\WaleckaMatrixEl_{LL} &= \big| \big< J_f \big\Vert
  \mathcal{\hat{L}}_J \big\Vert J_i \big> \big|^2
\\
\WaleckaMatrixEl_{T} &= \big| \big< J_f \big\Vert
  \mathcal{\hat{T}}^\mathrm{el}_J \big\Vert J_i \big> \big|^2
+ \big| \big< J_f \big\Vert
  \mathcal{\hat{T}}^\mathrm{mag}_J \big\Vert J_i \big> \big|^2
\\
\label{eq:last_Walecka_ME}
\WaleckaMatrixEl_{T^\prime} &= 2 \Re\!\Big[ \big< J_f \big\Vert
  \mathcal{\hat{T}}^\mathrm{el}_J \big\Vert J_i \big>
\big< J_f \big\Vert
  \mathcal{\hat{T}}^\mathrm{mag}_J \big\Vert J_i \big>^{*}
\Big] \,,
\end{align}
which are evaluated using a multipole expansion of the hadronic current. We
follow the usual treatment of these by evaluating the Coulomb
($\mathcal{\hat{M}}_J$), longitudinal ($\mathcal{\hat{L}}_J$), transverse
electric ($\mathcal{\hat{T}}^\mathrm{el}_J$) and transverse magnetic
($\mathcal{\hat{T}}^\mathrm{mag}_J$) operators to first order in inverse powers
of the nucleon mass $m_N$. Under this approach, we have~\cite{Serot:1978vj}
\begin{align}
\label{eq:first_Walecka_op}
\mathcal{\hat{M}}_J &= F_1\,M_J + \frac{ i\,\kappa }{ m_N }
\big[ F_A \, \Omega_J + \frac{1}{2} (F_A - \omega\,F_P)
\Sigma^{\prime\prime}_J \big] \\
\mathcal{\hat{L}}_J &= \frac{ \kappa }{ m_N }
\, \big( \Delta^{\prime\prime}_J
 -\frac{1}{2} \, M_J \big) F_1  - i\big( F_A
- \frac{ \kappa^2 }{ 2 m_N } \, F_P \big) \Sigma^{\prime\prime}_J \\
\mathcal{\hat{T}}^\mathrm{el}_J &= \frac{ \kappa }{ m_N } \big[
F_1 \, \Delta^{\prime}_J + \frac{1}{2} ( F_1 + 2 m_N F_2 )
\, \Sigma_J \big] - i F_A \, \Sigma^{\prime}_J
\\
\label{eq:last_Walecka_op}
\mathcal{\hat{T}}^\mathrm{mag}_J &=
- \frac{ i \kappa }{ m_N } \big[ F_1 \, \Delta_J - \frac{1}{2}
(F_1 + 2 m_N F_2 ) \, \Sigma^{\prime}_J \big] - F_A \, \Sigma_J
\end{align}
where expressions for the eight basic operators $M_J$, $\Delta_J$,
$\Delta^{\prime}_J$, $\Delta^{\prime\prime}_J$, $\Sigma_J$,
$\Sigma^{\prime}_J$, $\Sigma^{\prime\prime}_J$, and $\Omega_J$ are given in
Appendix~\ref{sec:basic_operators}.

\subsection{Nucleon form factors}
\label{sec:nucleon_form_factors}

The charged-current nucleon form factors $F_1$, $F_2$, $F_A$, and $F_P$ are
functions of the negative square of the four-momentum transfer
\begin{equation}
Q^2 \equiv -q^2 = -(\pnu - \plep)^2 = \kappa^2 - \omega^2 \,,
\end{equation}
or, equivalently, the variable
\begin{equation}
\tau \equiv \frac{ Q^2 }{ 4 \, m_N^2 } \,.
\end{equation}
To evaluate the Dirac form factor
\begin{equation}
\label{eq:Dirac_FF}
F_1 = \frac{ G_E^p - G_E^n + \tau\,G_M^p - \tau\,G_M^n }{ 1 + \tau } \,,
\end{equation}
and the Pauli form factor
\begin{equation}
\label{eq:Pauli_FF}
F_2 = \frac{ G_E^n - G_E^p - G_M^n + G_M^p }{ 2 \, (1 + \tau) \, m_N } \,,
\end{equation}
in terms of the proton and neutron Sachs form factors, we adopt the
BBBA05~\cite{Bradford:2006yz} parameterization
\begin{align}
G_E^p( \tau ) &= \frac{ 1 - 0.0578\,\tau }
  { 1 + 11.1\,\tau + 13.6\,\tau^2 + 33\,\tau^3 } \cdot g_V \\[1mm]
G_E^n( \tau ) &= \frac{ 1.25\,\tau + 1.30\,\tau^2}
  { 1 - 9.86\,\tau + 305\,\tau^2 - 758\,\tau^3 + 802\,\tau^4 } \\[1mm]
G_M^p( \tau ) &= \frac{ 1 + 0.150\,\tau }
  { 1 + 11.1\,\tau + 19.6\,\tau^2 + 7.54\,\tau^3 }
  \cdot \frac{ \mu_p }{ \mu_N } \\[1mm]
G_M^n( \tau ) &= \frac{ 1 + 1.81\,\tau }
  { 1 + 14.1\,\tau + 20.7\,\tau^2 + 68.7\,\tau^3 }
  \cdot \frac{ \mu_n }{ \mu_N }
\end{align}
where $g_V = 1$ is the nucleon vector coupling constant, $\mu_p$ ($\mu_n$) is
the proton (neutron) magnetic moment, and $\mu_N$ is the nuclear magneton. Note
that the expressions given in Eqs.~\ref{eq:Dirac_FF}~and~\ref{eq:Pauli_FF}
follow the convention that
\begin{equation}
F_1(0) + 2m_N F_2(0) = \frac{ \mu_p - \mu_n }{ \mu_N }
  \approx \num[round-mode = places, round-precision = 3]{4.70589} \,.
\end{equation}

We use a dipole parameterization for the axial-vector and pseudoscalar form
factors, i.e.,
\begin{align}
F_A &= F_A(Q^2) = \frac{ -g_A }{ \big( 1 + Q^2 / M_A^2 \big)^2 } \\[1mm]
F_P &= F_P( Q^2 ) = \frac{ 2 \, m_N \, F_A( Q^2 ) }{ m_\pi^2 + Q^2 }
\end{align}
where $g_A \approx 1.262$ is the nucleon axial-vector coupling constant, the
axial mass parameter is chosen to be $M_A = \SI{1032}{\MeV}$, and $m_\pi =
\SI[round-mode = places, round-precision = 2]{139.57039}{\MeV}$ is the mass of
a charged pion.

More sophisticated form factor parameterizations, such as those based on the
$z$ expansion formalism~\cite{Meyer:2016oeg, Borah:2020gte} are available in
the literature. However, at the tens-of-MeV neutrino energies relevant for this
study, the impact of different choices for the nucleon form factors is very
small.

\subsection{Differential cross section}
\label{sec:diff_xsec}

From the definitions above, one may write the inclusive differential cross
section for charged-current absorption as
\begin{align}
\nonumber \frac{ d^2\sigma }{ dT_\ell \, d\cos\theta_\ell }
 = & \; \frac{ G_F^2 \, |V_{ud}|^2 }
{ 2 \, \pi \, k \cdot p } \, \CoulombFactor \, E_\nu \, E_i \,
E_\ell \, |\mathbf{\plep}|
\\ \label{eq:diff_xsec_start} \times \;
& \delta\big( E_\nu + E_i - E_\ell - E_f \big) \cdot \RLep_{\mu \nu} \, \RNuc^{\mu \nu} \,,
\end{align}
where $T_\ell = E_\ell - m_\ell$ is the kinetic energy of the outgoing lepton.
The energy-conserving delta function is retained at this stage since it is
handled differently depending on whether the final nucleus is in a discrete
energy level or the unbound continuum.

\subsection{Discrete transitions}
\label{sec:discrete_xsec}

For a transition to a discrete energy level of the final nucleus, it is
convenient to specialize to the center-of-momentum (CM) frame, in which
Mandelstam $s = ( E_\nu + E_i )^2$ and
$\pnu \cdot \pNi = |\mathbf{\pnu}| \sqrt{s} \approx E_\nu \sqrt{s}$.
Integrating over $T_\ell$ eliminates the energy-conserving delta function
and yields the result
\begin{equation}
\label{eq:diff_xsec_discrete}
\frac{ d\sigma }{ d\cos\theta_\ell }
 = \frac{ G_F^2 \, |V_{ud}|^2 }
{ 2 \, \pi } \, \CoulombFactor \, \bigg[ \frac{ E_i \, E_f }{ s } \bigg]
E_\ell \, |\mathbf{\plep}|
\cdot \RLep_{\mu \nu} \, \RNuc^{\mu \nu} .
\end{equation}

\subsubsection{Allowed approximation}
\label{sec:allowed_approx}

In the previous version~\oldVersion\ of \marley~\cite{marleyPRC, marleyCPC}, the
allowed approximation (AA) was used to compute this differential cross section.
The AA combines the long-wavelength limit ($q \to 0$) and the slow-nucleon
limit ($|\mathbf{p}_N| \ll m_N$) to radically simplify the calculation. Under
the AA, the components of the weak hadronic charged current operator become
\begin{align}
j^\LorT &= g_V \, t_\mp & j^a &= -\sigma^a \, g_A \, t_\mp \,,
\end{align}
where $a\in\{\LorX,\LorY,\LorZ\}$. The upper (lower) sign on the isospin ladder
operator $t_\mp$ should be used for an incident neutrino (antineutrino).

The only non-vanishing elements of $\RNuc^{\mu \nu}$ under the AA are
\begin{align}
\RNuc^{\LorT \LorT} &= \FermiMat &
\RNuc^{aa} &= \frac{1}{3} \, \GTMat \,,
\end{align}
where
\begin{align}
\label{eq:BF}
\FermiMat &\equiv \frac{ g_V^2 }{ 2J_i + 1 } \Big|\big< J_f \, \big\lVert
\, \textstyle\sum_{\nucIndex=1}^A t_\mp(\nucIndex)
\, \big\rVert \, J_i \big> \Big|^2 \,
\intertext{is the Fermi matrix element, and}
\label{eq:BGT}
\GTMat &\equiv \frac{ g_A^2 }{ 2J_i + 1 } \Big|\big< J_f \, \big\lVert
\, \textstyle\sum_{\nucIndex=1}^A \boldsymbol{\sigma}(\nucIndex)
\, t_\mp(\nucIndex) \, \big\rVert \, J_i \big> \Big|^2 \,
\end{align}
is the Gamow-Teller (GT) matrix element. The tensor contraction needed to evaluate
the differential cross section in Eq.~\ref{eq:diff_xsec_discrete} then becomes
\begin{align}
\nonumber
\RLep_{\mu \nu} \, \RNuc^{\mu \nu}
= \Big(1 \, + & \, \beta_\ell \cos \theta_{\ell} \Big)\,
\FermiMat \\
\label{eq:AA_tensor_contraction}
& +\, \Big(1 - \frac{1}{3} \, \beta_\ell \cos \theta_{\ell} \Big)
\, \GTMat \,.
\end{align}
The two allowed matrix elements obey the parity selection rule
\begin{equation}
\label{eq:parity_selection_rule}
\Pi_f = \Pi_i \,,
\end{equation}
where $\Pi_i$ ($\Pi_f$) is the initial (final) nuclear parity. They also
obey the spin selection rules
\begin{align}
\FermiMat &= 0 \text{ unless } J_f = J_i \,, \\
\GTMat &= 0 \text{ unless } |J_i - 1| \leq J_f \leq J_i + 1 \,.
\end{align}

\subsubsection{Limitations of the allowed approximation}

As discussed in Ref.~\cite{marleyPRC}, multiple experimental
measurements~\cite{Bhattacharya1998, Bhattacharya2009, Liu1998} are available
at low $\isotope[40]{K}^{*}$ excitation energies for the allowed matrix
elements relevant for charged-current neutrino scattering on \isotope[40]{Ar}.
An advantage of the AA treatment described above is that the measured
$\FermiMat$ and $\GTMat$ values may be used directly, thus providing a
data-driven cross-section prediction for transitions to low-lying nuclear
energy levels.

However, this direct connection to data comes with two notable deficiencies
that we seek to mitigate in this article. First, the AA calculation continues
to use a discrete level treatment up to high excitation energies, localizing
the transition strength at individual points rather than providing the broad
distributions expected from nuclear giant resonances. Second, the AA by
definition neglects so-called \textit{forbidden} contributions to the cross
section, which arise from portions of the nuclear matrix element $\NucMat^\mu$
that vanish under the long-wavelength and slow-nucleon limits. As shown
specifically for an argon target in Refs.~\cite{VanDessel:2019atx,
VanDessel:2019obk}, the forbidden contributions become important at neutrino
energies of several tens of \si{\MeV}, affecting the total cross section as
well as differential energy and angle spectra for the outgoing lepton.

To address these deficiencies in the present work, \marley~\version\ implements
a HF-CRPA model of the neutrino-argon charged-current cross section, which we
use as a complete replacement for the previous AA calculation at high
$\isotope[40]{K}^{*}$ excitation energies. Details of the HF-CRPA model are
discussed below in Sec.~\ref{sec:continuum_xsec}.

Because the HF-CRPA approach is appropriate only within the unbound continuum,
a complete calculation of the inclusive cross section still requires a separate
treatment of transitions to discrete $\isotope[40]{K}^{*}$ nuclear levels.
While simply including the original \marley~AA model prediction for the needed
transitions might be reasonable as a first approximation, such a strategy leads
to immediate inconsistencies since the HF-CRPA model invokes neither of the two
limits that define the AA. In particular, since the allowed transitions are
generated by multipole operators proportional to the zeroth spherical Bessel
function of the first kind $j_0( \kappa \, r_n )$, which is equal to one for
$\kappa = 0$ and less than one for a nonzero argument, the AA will typically
overestimate the allowed strength whenever the momentum transfer becomes
appreciable.

To remove this specific limitation while preserving the connection to measured
$\FermiMat$ and $\GTMat$ values from the original \marley\ model, we develop an
improved treatment of the allowed discrete transitions that includes
approximate corrections for the momentum transfer dependence. Inclusion of
forbidden discrete transitions, which are expected to make a small contribution
to the inclusive cross section and have not yet been measured for argon, is
reserved for future work.

\subsubsection{Adjustments for nonzero momentum transfer}

The starting point for the updated model of discrete transitions is the set of
expressions involving eight basic operators in
Eqs.~\ref{eq:first_Walecka_op}--\ref{eq:last_Walecka_op}, which are obtained
from a multipole expansion of the nuclear current that keeps terms to first
order in $1/m_N$.
As shown in Appendix~\ref{sec:basic_operators}, all of the basic operators may
be written in terms of the quantity $M_J$. For a target nucleus with a
ground-state spin-parity of $0^+$ (such as \isotope[40]{Ar}), this choice of
notation allows one to identify the operators that generate allowed transitions
based on two criteria. First, the operator must contain the quantity $M_0$,
which itself contains the spherical Bessel function $j_0( \kappa \, r_n )$
mentioned previously. Second, to avoid inducing a change to the nuclear parity
that would violate the selection rule from Eq.~\ref{eq:parity_selection_rule},
the operator must have even parity.

Table~\ref{tab:operator_parities} summarizes the information needed to identify
the relevant operators. Each of the eight basic operators is listed together
with its parity as a function of the multipole order $J$. The third column
gives the multipole order $J_\text{allowed}$ for which the operator of interest
contains $M_0$. The final column lists the parity $\Pi_\text{allowed}$ of the
operator evaluated for $J = J_\text{allowed}$. For completeness, parities are
also given in the final two rows of the table for two spherical tensor products
that appear in the appendix. Because the relationship between the parameter $L$
and the multipole order $J$ is undefined in this context, the values of
$J_\text{allowed}$ and $\Pi_\text{allowed}$ are omitted for these.

\begin{table}
\renewcommand{\arraystretch}{1.4}
\caption{Properties of the basic operators.} \vspace{-0.8\baselineskip}
\begin{tabular}{c@{\hspace{3mm}}lc@{\hspace{3mm}}c}
\toprule
Operator & Parity & $J_\text{allowed}$ & $\Pi_\text{allowed}$ \\[-0.7mm]
\midrule
$M_J$ & $(-1)^{J}$ & $0$ & $+$ \\
$\Delta_J$ & $(-1)^{J+1}$ & $0$ & $-$ \\
${\Delta^\prime}_J$ & $(-1)^{J}$ & $1$ & $-$ \\
${\Delta^{\prime\prime}}_J$ & $(-1)^{J}$ & $1$ & $-$ \\
$\Sigma_J$ & $(-1)^{J}$ & $0$ & $+$ \\
${\Sigma^\prime}_J$ & $(-1)^{J+1}$ & $1$ & $+$ \\
${\Sigma^{\prime\prime}}_J$ & $(-1)^{J+1}$ & $1$ & $+$ \\
$\Omega_J$ & $(-1)^{J+1}$ & $0$ & $-$ \\
$\big[ M_L \otimes \boldsymbol{\sigma} \big]_{J}$
  & $(-1)^{L}$ & $^*$ & $^*$\\
$\big[ M_L \otimes \mathbf{p} \big]_{J}$
  & $(-1)^{L+1}$ & $^*$ & $^*$ \\
\bottomrule
\end{tabular}
\label{tab:operator_parities}
\end{table}

As shown in the table, four of the basic operators have odd parity for $J =
J_\text{allowed}$. These therefore do not contribute to the allowed
transitions, and we may ignore them in this context:
\begin{align}
\Delta_J &\to 0 &
\Delta^\prime_J &\to 0 &
\Delta^{\prime\prime}_J &\to 0 &
\Omega_J &\to 0 \,.
\end{align}
Since $\Sigma_J = 0$ for $J = J_\text{allowed} = 0$, only the remaining three
basic operators generate allowed transitions. Ignoring irrelevant terms that do
not contain $M_0$, we may write the three operators in the form
\begin{align}
M_0 = \frac{1}{ \sqrt{4\pi} } &\sum_{\nucIndex = 1}^A j_0( \kappa \, r_n )
  \, t_{\mp}(\nucIndex) \\[1mm]
\Sigma^\prime_1 = \sqrt{\frac{2}{3}} \cdot \frac{1}{ \sqrt{4\pi} }
  &\sum_{\nucIndex = 1}^A j_0( \kappa \, r_n )
  \, \boldsymbol{\sigma}(\nucIndex) \, t_{\mp}(\nucIndex) \\[1mm]
\Sigma^{\prime\prime}_1 = \frac{1}{\sqrt{3}} \cdot \frac{1}{ \sqrt{4\pi} }
  &\sum_{\nucIndex = 1}^A j_0( \kappa \, r_n )
  \, \boldsymbol{\sigma}(\nucIndex) \, t_{\mp}(\nucIndex) \,.
\end{align}
Evaluating the nuclear responses in terms of these operators leads to the
expressions
\begin{equation}
\NucResp_{CC} = \frac{ F_1^2 \, \mathcal{B}_{F}(\kappa) }{ g_V^2 }
  + \frac{ \kappa^2 }{ 4 \, m_N^2 } ( F_A + \omega \, F_P )^2 \,
\frac{ \mathcal{B}_{GT}(\kappa) }{ 3\, g_A^2 }
\end{equation}
\begin{equation}
\NucResp_{LL} = \frac{ \omega^2 \, F_1 ^2 \, \mathcal{B}_{F}(\kappa) }
  { \kappa^2 \, g_V^2 }
  + \bigg( F_A - \frac{ \kappa^2 }{ 2 \, m_N } \, F_P \bigg)^2
  \, \frac{ \mathcal{B}_{GT}(\kappa) }{ 3\, g_A^2 }
\end{equation}
\begin{equation}
\NucResp_{CL} = - \frac{ 2 \, \omega \, F_1^2
   \, \mathcal{B}_{F}(\kappa) }{ \kappa \, g_V^2 }
  - X_{AP} \, \frac{ \mathcal{B}_{GT}(\kappa) }{ 3\, g_A^2 }
\end{equation}
\begin{equation}
\NucResp_{T} = \bigg( F_A^2 + \frac{ \kappa^2 }{ 4 \, m_N^2 }
 \, \big[ F_1 + 2 \, m_N \, F_2 \big]^2 \bigg) \cdot
   \, \frac{ 2 \, \mathcal{B}_{GT}(\kappa) }{ 3 \, g_A^2 }
\end{equation}
\begin{equation}
\NucResp_{T^\prime} = - \frac{ \kappa }{ m_N } \, F_A
  \, \big( F_1 + 2 \, m_N \, F_2 \big) \cdot
   \frac{ 2 \, \mathcal{B}_{GT}(\kappa) }{ 3 \, g_A^2 }
\end{equation}
where we have defined the abbreviation
\begin{equation}
X_{AP} \equiv
  \frac{ ( F_A + \omega \, F_P ) \kappa }{ 2 \, m_N } \cdot
  \bigg( F_A - \frac{ \kappa^2 }{ 2 \, m_N } \, F_P \bigg) \,.
\end{equation}
Here the analogues of the Fermi and GT matrix elements with the
$\kappa$ dependence retained are denoted by the symbols
\begin{equation}
\label{eq:pseudo_BF}
\mathcal{B}_{F}(\kappa) \equiv \frac{ g_V^2 }{ 2J_i + 1 } \Big|\big< J_f
\, \big\lVert \, \sum_{\nucIndex=1}^A j_0(\kappa r_\nucIndex)
\, t_\mp(\nucIndex) \, \big\rVert \, J_i \big> \Big|^2 \,,
\end{equation}
and
\begin{equation}
\label{eq:pseudo_BGT}
\mathcal{B}_{GT}(\kappa) \equiv \frac{ g_A^2 }{ 2J_i + 1 } \Big|\big< J_f
\, \big\lVert \, \sum_{\nucIndex=1}^A j_0(\kappa r_n) \,
 \boldsymbol{\sigma}(\nucIndex)
\, t_\mp(\nucIndex) \, \big\rVert \, J_i \big> \Big|^2 \,,
\end{equation}
respectively.

If we define the additional abbreviations
\begin{align}
F_{PA} &\equiv \frac{ F_P }{ F_A } \\
\intertext{and}
F_{12A} &\equiv \frac{ F_1 + 2 \, m_N \, F_2 }{ F_A } \,,
\end{align}
then we may use the results above to write the tensor contraction from
Eq.~\ref{eq:diff_xsec_discrete} in the form
\begin{align}
\nonumber
\RLep_{\mu \nu} \, & \RNuc^{\mu \nu} =
\frac{ F_1^2 \, \mathcal{B}_F(\kappa) }{ g_V^2 }
\Bigg[ v_{CC} + \frac{ \kappa }{ m_N } \, v_{CL}
  + \frac{ \kappa^2 }{ 4 \, m_N^2 } \, v_{LL} \Bigg]
\\
\nonumber
& + \frac{ F_A^2 \, \mathcal{B}_{GT}(\kappa) }{ g_A^2 }
\Bigg[ \frac{\kappa ^2}{12 \, m_N^2} \left(
  1 - 2 \, \omega \, F_{PA}
  + \omega^2 \, F_{PA}^2 \right) v_{CC}
\\
\nonumber
& + \frac{ \kappa }{3 m_N} \left(1
- \big[ \omega + \frac{ \kappa^2 }{ 2 m_N } \big] \, F_{PA}
  + \frac{\omega \, \kappa^2 }{ 2 \, m_N } \, F_{PA}^2 \right) v_{CL}
\\
\nonumber
& + \left( \frac{1}{3} - \frac{ \kappa^2}{3 \, m_N} \, F_{PA}
  + \frac{ \kappa^4 }{ 12 \, m_N^2 } \, F_{PA}^2 \right) v_{LL}
\\
\label{eq:walecka_style_contraction}
& + \left( \frac{2}{3} + \frac{ \kappa^2 }{6 \, m_N^2} \, F_{12A}^2
\right) v_{T}
  - \chirality \, \frac{2 \kappa }{ 3 m_N } \, F_{12A} \, v_{T^\prime} \Bigg] \,,
\end{align}
where the $\LepFunc_\IndexForWalecka$ factors remain the same as in
Eqs.~\ref{eq:first_lep_func}--\ref{eq:last_lep_func}.

\subsubsection{Nuclear form factor}
\label{sec:nuclear_form_factor}

While the updated expression for the tensor contraction avoids the need to
invoke the long wavelength limit employed in the AA version from
Eq.~\ref{eq:AA_tensor_contraction}, the measured matrix elements $\FermiMat$
and $\GTMat$ can only be used to compute the right-hand side of
Eq.~\ref{eq:walecka_style_contraction} if a connection is established between
them and $\mathcal{B}_F(\kappa)$ and $\mathcal{B}_{GT}(\kappa)$ for nonzero
momentum transfers $\kappa$. Although rigorously evaluating this connection
will in general require a complicated many-body nuclear model calculation, we
resort here to a rough phenomenological approach that should nevertheless
represent a significant improvement over the AA previously used in \marley.

To derive the required approximate relationship between the two sets of allowed
matrix elements, it is helpful to recast
Eqs.~\ref{eq:pseudo_BF}~and~\ref{eq:pseudo_BGT} as
\begin{equation}
\label{eq:pseudo_BF_density}
\mathcal{B}_{F}(\kappa) = \frac{ g_V^2 }{ 2J_i + 1 }
  \sum_{M_i, M_f} \Big| \int r^2 j_0(\kappa r )
  \, \rho^{F}_{fi}(r) \, dr \Big|^2 \,,
\end{equation}
and
\begin{equation}
\label{eq:pseudo_BGT_density}
\mathcal{B}_{GT}(\kappa) = \frac{ 3\,g_A^2 }{ 2J_i + 1 }
  \sum_{M_i,M_f} \Big| \int r^2 j_0(\kappa r )
  \, \rho^{GT}_{fi}(r) \, dr \Big|^2 \,.
\end{equation}
Here the radial transition densities $\rho^{F}_{fi}(r)$ and $\rho^{GT}_{fi}(r)$
are given in terms of the initial ($\Psi_i$) and final ($\Psi_f$) nuclear
wave functions by
\begin{equation}
\rho^{F}_{fi}(r) \equiv
\sum_{\nucIndex = 1}^A \int \Psi_f^*( \mathbf{r} ) \, t_\mp(\nucIndex)
  \, \delta^{(3)}(\mathbf{r} - \mathbf{r}_n )
  \, \Psi_i( \mathbf{r} ) \, d^2\hat{\mathbf{r}} \,,
\end{equation}
and
\begin{equation}
\label{eq:GT_transition_density_2}
\rho^{GT}_{fi}(r) \equiv
\sum_{\nucIndex = 1}^A \int \Psi_f^*( \mathbf{r} ) \, \sigma_z(\nucIndex)
  \, t_\mp(\nucIndex) \, \delta^{(3)}(\mathbf{r} - \mathbf{r}_\nucIndex )
  \, \Psi_i( \mathbf{r} ) \, d^2\hat{\mathbf{r}} \,,
\end{equation}
where $\mathbf{r}_\nucIndex$ is the position of the $\nucIndex$-th nucleon and
$d^2\hat{\mathbf{r}}$ is the infinitesimal solid angle element.\footnote{One
obtains the same $\mathcal{B}_{GT}$ value when any of the three spherical
components of the Pauli vector $\sigma_{-1}$, $\sigma_{0}$, and $\sigma_{+1}$
is used to define the transition density in
Eq.~\ref{eq:GT_transition_density_2}. Without loss of generality, we choose
$\sigma_{z} = \sigma_{0}$.}

By setting $\kappa = 0$ in
Eqs.~\ref{eq:pseudo_BF_density}~and~\ref{eq:pseudo_BGT_density}, one may
immediately obtain the normalizations of the transition densities in terms of
the measured Fermi and GT strengths:
\begin{equation}
\FermiMat = \frac{ g_V^2 }{ 2J_i + 1 }
\sum_{M_i} \sum_{M_f}
\Big| \int r^2 \, \rho^{F}_{fi}(r) \, dr \Big|^2 \,,
\end{equation}
\begin{equation}
\GTMat = \frac{ 3\,g_A^2 }{ 2J_i + 1 }
\sum_{M_i} \sum_{M_f}
\Big| \int r^2 \, \rho^{GT}_{fi}(r) \, dr \Big|^2 \,.
\end{equation}

Using an approach similar to the one from Sec.~VIII of
Ref.~\cite{Schiff:1954zz}, we will make the approximation that the radial
dependence of the transition densities $\rho^{F}_{fi}$ and $\rho^{GT}_{fi}$ is
the same as the nucleon number density of the initial nucleus
\begin{equation}
\rho_i(r) =
\frac{1}{A} \sum_{\nucIndex = 1}^A \int \Psi_i^*( \mathbf{r} )
  \, \delta^{(3)}(\mathbf{r} - \mathbf{r}_\nucIndex )
  \, \Psi_i( \mathbf{r} ) \, d^2\hat{\mathbf{r}} \,,
\end{equation}
where we have chosen the normalization so that
\begin{equation}
\int r^2 \rho_i(r) \, dr = 1\,.
\end{equation}
Under this approximation, the nuclear matrix elements become
\begin{align}
\label{eq:B_with_F}
\mathcal{B}_{F}(\kappa) &\approx \FermiMat \, F^2(\kappa) \;\;
& \mathcal{B}_{GT}(\kappa) &\approx \GTMat \, F^2(\kappa) \,,
\end{align}
where
\begin{equation}
F(\kappa) = \int r^2 j_0(\kappa r) \, \rho_i(r) \, dr
\end{equation}
is the nuclear form factor commonly used in the literature for coherent elastic
neutrino-nucleus scattering (CEvNS). To evaluate this form factor, we adopt
the Klein-Nystrand expression~\cite{Klein:1999qj}
\begin{equation}
F(\kappa) = \frac{ 3 \, j_1( \kappa R_A ) }{ (1 + \kappa^2 a_k^2 ) \kappa R_A }
\end{equation}
with parameter values chosen as in the \textit{adapted} version from
Ref.~\cite{VanDessel:2020epd}. Namely, these are the Yukawa potential range
\begin{equation}
a_k = \SI{0.7}{\femto\meter}\,,
\end{equation}
and the effective nuclear radius
\begin{equation}
R_A = \sqrt{ \frac{5}{3}\,r_0^2 - 10\,a_k^2 } \,,
\end{equation}
where $r_0 = \SI{3.4274}{\femto\meter}$ is the measured~\cite{Angeli2013}
root-mean-square nuclear charge radius of \isotope[40]{Ar}.

\subsubsection{Summary of the discrete transition model}

To simulate transitions to discrete nuclear energy levels corresponding to
bound states, \marley~\version\ uses the expressions from
Eqs.~\ref{eq:walecka_style_contraction}~and~\ref{eq:B_with_F} when computing
the differential cross section in Eq.~\ref{eq:diff_xsec_discrete}. For the
results shown herein, we adopt values of $\FermiMat$ and $\GTMat$ that are
unchanged from the \texttt{ve40ArCC\_Bhattacharya1998.react} input file used
previously in Ref.~\cite{marleyPRC}, except that only nuclear energy levels
below the $\alpha$ particle emission threshold of \isotope[40]{K} are included.
All of the tabulated Fermi and GT strengths for the excitation energy range of
interest are taken directly from the $\beta^{+}$~decay data set reported in
Ref.~\cite{Bhattacharya1998}.

\subsection{Continuum transitions}
\label{sec:continuum_xsec}

For transitions to final excitation energies in the unbound continuum, the
previous version \oldVersion\ of \marley\ maintained the AA treatment presented
in Sec.~\ref{sec:allowed_approx} above. The continuum was approximated using a
set of discrete energy levels and corresponding theoretical GT strengths taken
from Ref.~\cite{Cheoun2012a}. In the present version~\version, this approach has
been entirely replaced as described below.

Starting from the differential cross section in Eq.~\ref{eq:diff_xsec_start},
one may specialize to the laboratory frame rather than the CM frame used in
Sec.~\ref{sec:discrete_xsec}. In the laboratory frame, $\pnu \cdot \pNi = E_\nu
\, E_i$. If the definitions of the nuclear responses $\NucResp_\IndexForWalecka$
from Eq.~\ref{eq:NucResponseFunc} are adjusted to include the energy-conserving
delta function~\cite{Walecka1975}, one obtains the expression
\begin{equation}\label{eq:diff_xsec_continuum}
\frac{ d^2\sigma }{ dT_\ell \, d\cos\theta_\ell }
 = \frac{ G_F^2 \, |V_{ud}|^2 }{ 2 \, \pi } \, \CoulombFactor
\, E_\ell \, |\mathbf{\plep}| \cdot \RLep_{\mu \nu} \, \RNuc^{\mu \nu} \,.
\end{equation}
All multipoles corresponding to $0 \leq J \leq 5$ and both parities are included
in the new \marley\ calculation of the continuum differential cross section. The
response functions $\NucResp_\IndexForWalecka$ in the continuum are evaluated
according to an HF-CRPA approach. Here, we briefly outline the essential aspects
of this approach while referring readers to Refs.~\cite{Jachowicz:1998fn,
Jachowicz:2002rr, Pandey:2014tza, Pandey:2016jju, VanDessel:2019atx,
VanDessel:2019obk} for a detailed discussion of the formalism.

\subsubsection{HF-CRPA nuclear responses}

The model begins with an HF description of the nucleus, solving the
Schr\"{o}dinger equation for nucleons in a self-consistent nuclear potential
derived using the Skyrme (SkE2) nucleon-nucleon interaction. The SkE2
interaction is parameterized to reproduce static properties, such as charge
radii and excitation energies, of closed-shell nuclei~\cite{Waroquier:1986mj}.
The HF framework provides a mean-field description where nucleons interact with
the averaged field generated by all other nucleons, incorporating correlations
captured in the HF propagator. This results in a Slater determinant
representation of the nucleus, with nucleons occupying well-defined angular
momentum and energy states. Since the wave functions for the outgoing nucleon
are obtained by solving the positive-energy Schr\"{o}dinger equation with the
same nuclear potential used for the ground-state calculation, the effects of
elastic final-state interactions (FSI) are inherently incorporated. The approach
also naturally accounts for Fermi motion, Pauli blocking, and the binding energy
for the different bound nucleon levels. The HF procedure generates both
bound-state and continuum wave functions within the same approach, which ensure
orthogonality between initial and final nuclear mean-field states.

Long-range correlations between the nucleons are then introduced via a CRPA
treatment, which uses the same SkE2 interaction parameterization adopted for the
mean-field potential as the residual interaction. The CRPA formalism extends the
HF approach by describing nuclear excitations as coherent superpositions of
particle-hole ($ph^{-1}$) and hole-particle ($hp^{-1}$) states out of a
correlated ground state
\begin{equation}
| \Psi^C_{\rm{CRPA}} \rangle = \sum_{C'} \left\lbrace X_{C,C'}
  |p'h'^{-1}\rangle - Y_{C,C'} |h'p'^{-1}\rangle \right\rbrace,
\end{equation}
where the summation index $C$ denotes a set of quantum numbers defining an
excitation channel:
\begin{equation}
C = \left\lbrace n_h,l_h,j_h,m_{j_h},\varepsilon_h;l_p,j_p,m_{j_p},\tau_z
  \right\rbrace.
\end{equation}
The indices $p$ and $h$ represent the quantum numbers related to the particle
or the hole state, $\varepsilon_h$ denotes the binding energy of the hole
state, and $\tau_z$ defines the isospin character of the particle-hole pair.
Since the CRPA approach describes nuclear excitations as the coherent
superposition of individual particle-hole states out of a correlated ground
state, it enables the description of collective effects in the nucleus.
The response of the nucleus to external probes that enter in
Eq.~\ref{eq:diff_xsec_continuum} is encapsulated in the CRPA polarization
propagator, $\Pi^{\rm{CRPA}}(x_1,x_2,E_x)$, solved in coordinate space. This
propagator is derived from the iterative equation:
\begin{equation}\label{eq:crpa-propogator}
\begin{aligned}
&\Pi^{\rm{CRPA}}(x_1,x_2,E_x) = \Pi^{(0)}(x_1,x_2,E_x) \\
&+ \frac{1}{\hbar} \int \mathrm{d}x \int \mathrm{d}x'
  \left[ \Pi^{(0)}(x_1,x,E_x) \right. \\ &\times \left.
  \tilde{V}(x,x') \, \Pi^{\rm{CRPA}}(x',x_2,E_x) \right],
\end{aligned}
\end{equation}
where $\Pi^{(0)}$ represents the HF contribution, $\tilde{V}(x,x')$ is the
antisymmetrized SkE2 residual interaction, $E_x$ is the excitation energy of
the outgoing nucleus, and $x$ is the shorthand notation for the combination of
the spatial, spin, and isospin coordinates. Notably, while the HF-CRPA model is
particularly effective in describing reactions within the low-energy region
discussed here, it has also demonstrated success in addressing quasielastic
scattering at medium energies relevant to accelerator-based neutrino
probes~\cite{Pandey:2013cca, Pandey:2014tza, Pandey:2016jju,
VanDessel:2017ery}.

\subsubsection{Folding procedure}
\label{sec:lorentzian_folding}

A limitation of the CRPA formalism is its restriction of the configuration space
to one-particle one-hole excitations out of a correlated ground state.
Consequently, only the escape-width contribution to the final-state interaction
is accounted for, while the width of the initial single-nucleon states is not
taken into account. This limits the description of the transition to the
continuum states in the CRPA formalism. Although the energy location of excited
states is typically well predicted, the width tends to be  underestimated, and
the height of the response at the peak is overestimated. To address this, we
adopt a simplified phenomenological approach where modified response functions
$\NucRespWide_\IndexForWalecka$ are obtained by folding the HF and CRPA response
functions $\NucResp_\IndexForWalecka = \NucResp_\IndexForWalecka( \omega,
\kappa)$ with a Lorentzian distribution~\cite{Pandey:2014tza}:
\begin{equation}
\label{folding}
  \NucRespWide_\IndexForWalecka( \omega, \kappa )
  \equiv \int_{\omega - 4\,\LorentzianWidth}^{\omega + 4\,\LorentzianWidth}
  d\omegaWide \, \NucResp_\IndexForWalecka( \omegaWide, \kappa )
  \, L( \omega, \omegaWide ) \,,
\end{equation}
where
\begin{equation}
  L( \omega, \omegaWide ) = \LorentzianNorm
  \bigg[ \frac{\LorentzianWidth}
    { (\omega - \omegaWide)^2 + (\LorentzianWidth / 2)^2 }
  \bigg] \,,
\end{equation}
and the normalization constant $\LorentzianNorm$ is chosen so that
\begin{equation}
  \int_{\omega - 4\,\LorentzianWidth}^{\omega + 4\,\LorentzianWidth}
  d\omegaWide \, L( \omega, \omegaWide ) = 1 \,.
\end{equation}
We use an effective width of $\LorentzianWidth$ = 3 MeV which aligns well with
the predicted energy widths in the giant-resonance region. We also adopt the
notation $\RNucWide^{\mu\nu}$ to represent the Lorentzian-folded version of
the hadronic tensor. The elements of this tensor are evaluated using the
modified response functions from Eq.~\ref{folding} above, e.g.,
\begin{equation}
\RNucWide^{\LorT \LorT} = \NucRespWide_{CC} \,.
\end{equation}

This folding procedure redistributes strength from the peak to the tails of the
response, effectively accounting for the finite width of single-particle
excitations in an approximate manner. Importantly, it does not affect the
integrated cross section strength. However, folding introduces a new challenge:
HF-CRPA contributions to the cross section can spread to unphysical regions of
energy transfer above the $\omega = \kappa$ bound and below the continuum
threshold. Due to the rapid decrease of the cross section at high energy
transfer, the strength for $\omega > \kappa$ is negligible. Kinematic cuts
imposed by \marley\ during event generation ensure that this region is never
populated.

\subsubsection{Reassignment of sub-continuum strength}
\label{sec:sub_continuum}

As described in Sec.~\ref{sec:deex} below, the \marley\ de-excitation model
considers emission of $\gamma$-rays and nuclear fragments with nucleon number
$A \leq 4$. Therefore, \marley\ defines the excitation energy threshold for the
continuum
\begin{equation}
  \label{eq:continuum_threshold}
  \ExContinuumThresh \equiv
  \min_{ \FragIdx \in \{ n, p, d, t, h, \alpha \} } S_\FragIdx
\end{equation}
as the minimum of the separation energies
\begin{equation}
\label{eq:sep_energy}
  S_\FragIdx = m_\FragIdx
  + \mNuclear( Z^\prime - Z_\FragIdx, A - A_\FragIdx )
  - \mNuclear( Z^\prime, A )
\end{equation}
for all fragments that may be emitted from the outgoing nucleus, which has
proton (nucleon) number $Z^\prime$ ($A$) before the fragment is removed. The
fragment itself has proton (nucleon) number $Z_\FragIdx$ ($A_\FragIdx$) and
mass $m_\FragIdx$. The nuclear masses $\mNuclear$ are functions of both proton
and nucleon number. If one neglects electron binding energy (as is done
throughout \marley), then Eq.~\ref{eq:sep_energy} may be rewritten in terms of
atomic masses as
\begin{equation}
  S_\FragIdx \approx \mathpzc{M}_{\FragIdx}
  + \mAtom( Z^\prime - Z_\FragIdx, A - A_\FragIdx )
  - \mAtom( Z^\prime, A )
\end{equation}
where
\begin{equation}
  \mathpzc{M}_{\FragIdx} \equiv m_\FragIdx + Z_\FragIdx\,m_e
\end{equation}
and $m_e = \SI{0.51099893}{\MeV}$ is the electron mass.

In general, the redistribution of the cross section due to folding can lead to
significant strength below the continuum threshold \ExContinuumThresh,
especially for the $1^+$ multipole. To ensure compatibility with the nuclear
de-excitation model, \marley\ checks that the excitation energy of the outgoing
nucleus
\begin{equation}
  E_x = E_x( \omega, \kappa ) = \sqrt{ ( m_i + \omega )^2 - \kappa^2 }
    - m_f^\mathrm{g.s.} \,,
\end{equation}
is not less than \ExContinuumThresh\ in each simulated primary interaction.
Here $m_i$ is the mass of the target atom
\begin{equation}
  m_i = \mAtom( Z, A ) \,,
\end{equation}
which has proton (nucleon) number $Z$ ($A$). The symbol $m_f^\mathrm{g.s.}$
represents the ground-state mass of the outgoing ion, which (neglecting
electron binding energy) is given by
\begin{equation}
  m_f^\mathrm{g.s.} \approx \mAtom( Z^\prime, A ) + \chirality\,m_e \,.
\end{equation}

To handle cases where $E_x$ falls below the continuum threshold
$\ExContinuumThresh$, \marley\ treats the total energy of the outgoing lepton
as a function of excitation energy and scattering angle:
\begin{equation}
  \label{eq:El_as_function_of_Ex}
  E_\ell( E_x, \theta_\ell ) = \frac{ \sgn(\cos\theta_\ell)
  \, \big( \quadB^2 - 4\,\quadA\,\quadC \big)^{1/2} -\quadB }
  { 2 \, \quadA } \,,
\end{equation}
where the quadratic equation coefficients
\begin{align}
  \quadA &= 4 \, E_\mathrm{tot}^2 - 4 \, |\mathbf{\pnu}|^2
    \cos^2\theta_\ell \\
  \quadB &= 4 \, E_\mathrm{tot} \, \quadD^2 \\
  \quadC &= \quadD^4 + 4 \, m_\ell^2 \, |\mathbf{\pnu}|^2
    \cos^2\theta_\ell \,,
\end{align}
are defined in terms of the quantities
\begin{equation}
  \quadD^2 = ( m_f^\mathrm{g.s.} + E_x )^2
    - m_\ell^2 + |\mathbf{\pnu}|^2 - E_\mathrm{tot}^2 \,,
\end{equation}
and
\begin{equation}
  E_\mathrm{tot} = E_\nu + m_i \,.
\end{equation}
While holding the sampled value of $\theta_\ell$ constant, \marley\
recalculates $E_\ell$ by evaluating Eq.~\ref{eq:El_as_function_of_Ex} with $E_x
= \ExContinuumThresh$. All other kinematic variables in the event are then
updated accordingly. This procedure thus reassigns the sub-continuum portion of
the cross section to an excitation energy of exactly \ExContinuumThresh.

For the neutrino-argon results shown in this article, the continuum threshold
is the $\isotope[40]{K}^*$ separation energy for $\alpha$ particles:
\begin{equation}
\ExContinuumThresh = S_\alpha = \SI{6.438}{\MeV} \,.
\end{equation}
After applying a final adjustment for energy balance that we discuss below, the
impact of the $E_x \geq \ExContinuumThresh$ requirement in our model of the
$\isotope[40]{Ar}( \nu_e, e^- )\isotope[40]{K}^*$ reaction is small.

\subsubsection{Energy balance}
\label{sec:crpa_energy_balance}

As noted earlier, the HF-CRPA calculations presented herein use the same
potential to describe the nucleus in both the initial and final states.
However, the assumption of a consistent potential is violated because the CC
primary interaction changes the nuclear proton number $Z$ by one unit, i.e.,
\begin{equation}
 Z^\prime = Z - \chirality \,.
\end{equation}
To approximately correct for this, we introduce a parameter $\DeltaCRPA$, which
represents the energy difference between the ground state of the initial
nucleus and its isobaric analog state (IAS) in the outgoing nucleus. We then
define an effective value of the energy transfer $\omegaEff$ which is shifted
by this energy difference:
\begin{equation}
\omegaEff \equiv \omega + \DeltaCRPA \,.
\end{equation}
Since the Lorentzian-folded nuclear tensor is a function of the energy transfer
$\omega$ and the three-momentum transfer $\kappa$,
\begin{equation}
\RNucWide^{\mu \nu} = \RNucWide^{\mu \nu}( \omega, \kappa )\,,
\end{equation}
one may account for the energy difference between the initial and final nuclear
states by evaluating it using the effective energy transfer:
\begin{equation}
\label{eq:EffectiveRescaledNuclearTensor}
\RNucEff^{\mu \nu} \equiv \RNucWide^{\mu \nu}( \omegaEff, \kappa )\,.
\end{equation}
This leads to an updated expression for the differential cross section in the
continuum:
\begin{equation}
\label{eq:diff_xsec_continuum_DeltaCorrected}
\frac{ d^2\sigma }{ dT_\ell \, d\cos\theta_\ell }
 = \frac{ G_F^2 \, |V_{ud}|^2 }{ 2 \, \pi } \, \CoulombFactor
  \, E_\ell \, |\mathbf{\plep}| \cdot \RLep_{\mu \nu} \, \RNucEff^{\mu \nu} \,.
\end{equation}
Note that, apart from the substitution $\RNuc^{\mu \nu} \to \RNucEff^{\mu
\nu}$, all other quantities remain unchanged from
Eq.~\ref{eq:diff_xsec_continuum}. In particular, the kinematic factors that
appear in $\RLep_{\mu \nu}$ should be evaluated using $\omega$ rather than
$\omegaEff$.

For the neutrino-argon results shown later in this paper, we set $\DeltaCRPA =
\SI{-5.377}{\MeV}$. This value has two contributions that are summed together.
The first is the \SI{0.993}{\MeV} nuclear mass difference between the ground
states of \isotope[40]{K} and \isotope[40]{Ar}. The second is the
\SI{4.384}{\MeV} excitation energy~\cite{Chen:2017ngq} of the known isobaric
analog of the \isotope[40]{Ar} ground state in $\isotope[40]{K}^*$.

\subsubsection{Summary of the continuum transition model}

Above the excitation energy threshold $E_x^c$ defined in
Eq.~\ref{eq:sep_energy}, \marley\ models the structure of the outgoing nucleus
from the primary interaction using a continuous level density. Nuclear
transitions to the continuum are described using the differential cross section
from Eq.~\ref{eq:diff_xsec_continuum_DeltaCorrected}. The nuclear responses
corresponding to components of the hadronic tensor $\RNucEff^{\mu \nu}$ are
calculated via an HF-CRPA approach and then folded with a Lorentzian
distribution in order to approximately account for the width of the initial
single-nucleon states. To correct for the change in the nuclear potential that
occurs in the CC reaction, the folded nuclear responses are evaluated at an
effective value $\omegaEff$ of the energy transfer that is shifted from the
actual value. The size of the energy shift $\DeltaCRPA$ is equal to the energy
difference between the ground state of the target nucleus and its isobaric
analog state in the outgoing nucleus. To prevent problems in connecting this
treatment to the nuclear de-excitation model, any HF-CRPA cross-section
strength that falls below the continuum is reassigned to an excitation energy
of exactly $E_x^c$.

\subsection{Coulomb corrections}
\label{sec:coulomb_corrections}

In the CC reaction, the Coulomb potential of the residual nucleus distorts the
wave function of the outgoing lepton. This distortion has a significant impact
on the cross section at the low energies of interest for this work. To
approximately account for this effect, we apply a Coulomb correction factor
$\CoulombFactor$ in the calculation of the squared amplitude
$\SquaredAmplitude$ from Eq.~\ref{eq:CC_squared_amplitude}. Based on the
results from Ref.~\cite{Engel:1997fy} and the strategy used in similar neutrino
cross-section predictions~\cite{Volpe2002, SajjadAthar:2005ke, Ydrefors2012a,
VanDessel:2019atx, VanDessel:2019obk}, we evaluate $\CoulombFactor$ by
interpolating between two distinct expressions. Our approach is unchanged from
the one used in \marley~\oldVersion~\cite{marleyPRC}.

For low energies of the outgoing lepton, $\CoulombFactor$ is given by the Fermi
function~\cite{Fermi1934Original,Fermi1934Translation}
\begin{equation}
\label{eq:Fermi_function}
F_\mathrm{Fermi} = \frac{ 2 \, (1+S)
  \left|\Gamma\left( S + i\eta \right) \right|^2 e^{-\pi\,\eta} }
  { \big[\Gamma(1+2S)\big]^2 \, (2 \, \FNRpLep \, R)^{2-2S} }
  \,,
\end{equation}
commonly used in studies of $\beta$ decay. This function represents the ratio
of the s-wave solution for the lepton wave function in the Coulomb potential of
the nucleus to the plane-wave solution, both evaluated at the nuclear radius
\begin{equation}
\label{eq:nuclear_radius}
  R \approx \frac{ 1.2 \, A^{1/3} \, \text{fm} }{ \hbar\,c }\,.
\end{equation}
Here
\begin{equation}
\label{eq:Sommerfeld_param}
\eta = \frac{ \alpha \, Z^\prime \, \chirality }{ \beta_\mathrm{rel} }
\end{equation}
is the Sommerfeld parameter and
\begin{equation}
  S \equiv \sqrt{1 - \alpha^2 {Z^\prime}^2\,} \,,
\end{equation}
both of which are defined in terms of the fine structure constant $\alpha$.
The symbol $\FNRpLep$ denotes the momentum of the final lepton in the
rest frame of the outgoing ion. In terms of the
relative speed of these two particles~\cite{Cannoni:2016hro}
\begin{equation}
\label{eq:Lorentz_invariant_relative_speed}
  \beta_\mathrm{rel} = \frac{ \sqrt{ (\plep \cdot \pNf)^2
  - m_\ell^2 \, m_f^2 } }{ \plep \cdot \pNf } \,,
\end{equation}
and the Lorentz factor
\begin{equation}
  \gamma_\mathrm{rel} \equiv \left(1 - \beta^2_\mathrm{rel}\right)^{-1/2} \,,
\end{equation}
the total energy $\FNReLep$ and momentum $\FNRpLep$ of the lepton
in this frame may be written as
\begin{align}
\FNReLep &\equiv \gamma_\text{rel} \, m_\ell
& \FNRpLep &\equiv \beta_\text{rel} \, \FNReLep \,.
\end{align}
The mass of the outgoing ion $m_f$ includes its excitation energy, i.e.,
\begin{equation}
  m_f = E_x + m_f^\mathrm{g.s.} \,.
\end{equation}

At higher lepton energies, contributions beyond the s-wave portion of the wave
function become appreciable, and the Fermi function thus becomes a poor
approximation. For this energy region, the Modified Effective Momentum
Approximation (MEMA) developed in Ref.~\cite{Engel:1997fy} provides a suitable
alternative. Under the MEMA, the correction factor $\CoulombFactor$ is
evaluated as
\begin{equation}
\label{eq:FMEMA}
F_\mathrm{MEMA} \equiv \frac{ \FNRpLepEff \, \FNReLepEff }
{ \FNRpLep \, \FNReLep } \,.
\end{equation}
The effective values of the lepton energy and momentum used here are those
within the Coulomb potential at the center of the nucleus
\begin{align}
\label{eq:eff_vars}
\FNRpLepEff &\equiv \sqrt{ \FNReLepEff^2 - m_\ell^2 } &
\FNReLepEff &\equiv \FNReLep - V_C(0) \,,
\end{align}
which is taken to be that of a sphere with a uniform charge density:
\begin{equation}
V_C(0) \approx \frac{ 3 \, Z^\prime \, \chirality \, \alpha }{ 2 \, R } \,.
\end{equation}

We combine the Fermi function and MEMA in our calculations by choosing
the factor that leads to the smaller correction to the cross section:
\begin{equation}
\label{eq:CoulombFactor}
\CoulombFactor \equiv \begin{cases}
  F_\mathrm{Fermi} & |F_\mathrm{Fermi} - 1| < |F_\mathrm{MEMA} - 1| \\
  F_\mathrm{MEMA} & \text{otherwise}
\end{cases} \,.
\end{equation}
This interpolation scheme leads to $F_\mathrm{Fermi}$ being used at low lepton
energies and $F_\mathrm{MEMA}$ at high energies, with the transition between
the two taking place at approximately \SI{40}{\MeV} for a \isotope[40]{Ar}
target.

It should be noted that our approach to Coulomb corrections omits a second
adjustment recommended in the original MEMA~\cite{Engel:1997fy}: in addition to
an overall rescaling, the amplitude is also evaluated using the effective
lepton momentum instead of the unmodified momentum. When combining the full
MEMA with the Fermi function (which includes only rescaling), we find that the
second correction has a small overall impact on the present calculations. It
can also lead to pathologies in differential distributions, such as a sudden
shift in the derivative of the cross section at forward scattering angles. For
these reasons, we apply only the rescaling from Eq.~\ref{eq:CoulombFactor} in
all results shown in this article. Further study of Coulomb corrections in this
energy regime, including the (in)compatibility of the Fermi function and MEMA
approaches, would be a useful direction for future research.

\subsection{Nuclear de-excitations}
\label{sec:deex}

As described above, modeling of the primary neutrino interaction has been
extensively updated in the new version~\version\ of \marley. Conversely, the
approach to de-excitations of the outgoing nucleus has received only modest
adjustments since version~\oldVersion~\cite{marleyPRC}. For transitions to
discrete nuclear levels, $\gamma$-ray emission continues to be described using
the same tabulated branching ratios taken from version~1.6 of the TALYS nuclear
reaction code~\cite{TALYS16, Talys2023, Talys1}. In the unbound continuum, a
series of binary decays is simulated according to the Hauser-Feshbach
formalism~\cite{Hauser1952} and the procedures defined in the original \marley\
publications~\cite{marleyPRC, marleyCPC}. For the emission of a nuclear fragment
of type $\FragIdx$, where $\FragIdx$ may be a neutron, proton, or a light ion
($A \leq 4$), the differential decay width is computed as
\begin{equation}
\label{eq:fragment_diff_decay_width}
\frac{ d\Gamma_{\FragIdx} }{ dE_x^\prime }
= \frac{1}{2 \, \pi \, \rho_i }
\sum_{\ell = 0}^{ \ell_\mathrm{max} }
\;
\sum_{j = |\ell - s|}^{\ell + s}
\;
\sum_{J^\prime = |J - j|}^{J + j}
T_{\ell j} \, \rho_f \,,
\end{equation}
while for continuum $\gamma$-rays the expression
\begin{equation}
\label{eq:gamma_diff_decay_width}
\frac{ d\Gamma_\gamma }{ dE_x^\prime }
= \frac{1}{2 \, \pi \, \rho_i }
\sum_{\lambda = 1}^{ \lambda_\mathrm{max} } \;
\sum_{J^\prime = |J - \lambda|}^{J + \lambda} \;
\sum_{ \Pi^\prime \in \{-1, 1\} }
T_{X\lambda} \, \rho_f\,,
\end{equation}
is used. Here the initial (final) level density $\rho_i$ ($\rho_f$) is a
function of the initial (final) excitation energy $E_x$ ($E_x^\prime$), total
spin $J$ ($J^\prime$), and parity $\Pi$ ($\Pi^\prime$) of the nucleus and is
evaluated according to the back-shifted Fermi gas model~\cite{Koning2008}.
Contributions to the decay widths diminish as the fragment orbital angular
momentum $\ell$ and $\gamma$-ray multipolarity $\lambda$ grow, so the sums over
these quantities are truncated at $\ell_\mathrm{max} = \lambda_\mathrm{max} =
5$. The spin $s$ and total angular momentum $j$ of the emitted fragment appear
in two of the remaining sums, which together consider all allowed values of the
relevant observables.

The calculation of the $\gamma$-ray transmission coefficients $T_{X\lambda}$
uses the Standard Lorentzian parameterization from RIPL-3~\cite{RIPL3} and is
unchanged from \marley~\oldVersion. While the general approach to the fragment
transmission coefficients $T_{\ell j}$ has also been retained from the previous
code version, the algorithm used to compute them has been modified. In addition
to technical improvements related to numerical stability, execution speed, etc.,
the model of the nuclear optical potential used as input has been updated. In
\marley~\oldVersion, the original 46-parameter global optical potential
developed by Koning and Delaroche~\cite{Koning2003} was used for all fragment
transmission coefficient calculations. In the new version~\version\ of the code,
the same functional form is used for the optical potential, but all parameters
have been updated to match the \textit{KDUQFederal} recommendation reported in
Ref.~\cite{Pruitt2023}. This change was made to enable the impact of the optical
model parameter uncertainties, newly quantified for the Koning-Delaroche model
in this recent evaluation, to be studied for the first time in the context of
low-energy neutrino scattering. Details of the initial uncertainty studies will
be discussed in a forthcoming \marley\ publication.

For the present results, the most significant change to the de-excitation model
is related to handling of the input data rather than the decay width
calculations themselves. In both \marley\ versions, the code is accompanied by
data tables taken from TALYS that provide information about discrete nuclear
levels and associated $\gamma$-ray transitions. Although tabulated levels above
the continuum threshold $\ExContinuumThresh$ from
Eq.~\ref{eq:continuum_threshold} are allowed to appear in the nuclear structure
tables, these levels are consistently ignored by both versions of \marley\
during simulation of the primary neutrino interaction. In the case of
\marley~\oldVersion, which includes only discrete transitions in the primary
interaction, the excitation energy associated with each input Fermi or
Gamow-Teller matrix element is taken at face value above $\ExContinuumThresh$.
Below the continuum threshold, both code versions match the matrix element input
(based solely on closest excitation energy) to known nuclear levels from the
structure data tables.

Treatment of the continuum becomes different between code versions in the
de-excitation stage of the simulation. In \marley~\oldVersion, only excitation
energies above the last tabulated level for a given final-state nuclide are
considered to lie within the continuum. This definition is followed regardless
of the completeness of the level scheme or the value of $\ExContinuumThresh$
for the nucleus of interest. For nuclei with many measured energy levels, this
can lead to a much higher continuum threshold than would be obtained via
Eq.~\ref{eq:continuum_threshold}. In the new version of \marley, nuclear levels
above the calculated $\ExContinuumThresh$ are ignored for de-excitations as
well as the primary interaction. As will be shown below, this change
strongly enhances the cross section leading to a one-neutron one-proton
($1n1p$) final state. It also has a significant impact on the cross section for
other exclusive channels.

\section{Results}
\label{sec:results}

In this section, we present predictions of the updated \marley~2.0.0 neutrino
interaction model explained above. We denote this model by the abbreviation
\marleyNew\ in plot legends and descriptions. When \marley~\oldVersion\
predictions are also shown, they are denoted using \marleyOld.

\subsection{Impact of incremental model changes}

\begin{figure}
\includegraphics[width=0.5\textwidth]{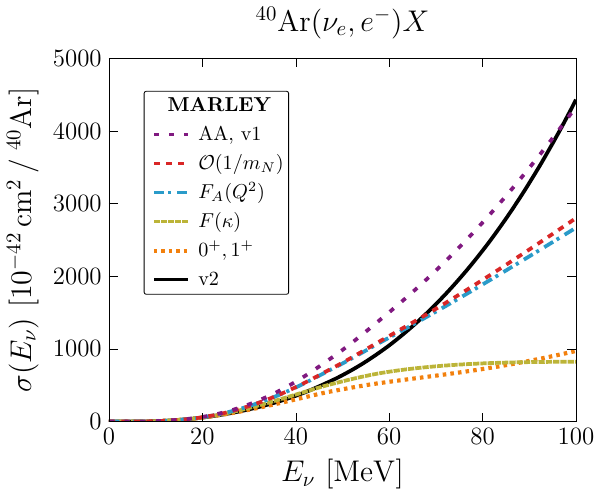}
\caption{Evolution of the \marley\ prediction of the inclusive
  $\nu_e$-\isotope[40]{Ar} charged-current total cross section. Corrections
  described in the text are incorporated progressively starting from the
  \marley~\oldVersion\ model (\marleyOld, loosely-dashed violet) until the full
  \marley~\version\ calculation is obtained (\marleyNew, solid black).}
\label{fig:model_evolution}
\end{figure}

Figure~\ref{fig:model_evolution} shows the evolution of the \marley\
calculation of the inclusive total cross section $\sigma(E_\nu)$ for
charged-current $\nu_e$ absorption on \isotope[40]{Ar}. The loosely-dashed
violet line gives the $\sigma(E_\nu)$ prediction under the AA defined in
Sec.~\ref{sec:allowed_approx}. The AA result is identical to the cross section
predicted by \marleyOld. To obtain the cross section shown by the dashed red
line, the AA tensor contraction from Eq.~\ref{eq:AA_tensor_contraction} is
replaced by the $\mathcal{O}(1/m_N)$ expression from
Eq.~\ref{eq:walecka_style_contraction}. However, the nucleon and nuclear form
factors are all still evaluated in the low-momentum transfer limit $q \to 0$.
The use of discrete nuclear levels at all excitation energies is also retained
from \marleyOld.

Introducing the full $Q^2$ dependence of the nucleon form factors into the
calculation yields the result shown by the dot-dashed cyan line, which is very
close to the dashed red one except at the highest neutrino energies plotted. In
contrast to this small correction, evaluating the nuclear form factor as a
function of $\kappa$ dramatically reduces the cross section above roughly
$E_\nu = \SI{40}{\MeV}$. The impact of this improvement to the calculation is
shown by the densely-dashed yellow line.

So far, all of the plotted cross sections have considered only nuclear
transitions to $0^{+}$ and $1^{+}$ final states. The matrix elements for all
other transitions vanish under the AA. In the calculation represented by the
dotted orange line, the discrete level treatment at excitation energies above
the continuum threshold $\ExContinuumThresh$ is replaced by the HF-CRPA model
described in Sec.~\ref{sec:continuum_xsec}, but only for the allowed multipoles
($0^{+}$ and $1^{+}$) considered in previous steps. Inclusion of the forbidden
multipoles (all others of both parities up to order $J = 5$) leads to the full
\marleyNew\ result, which is plotted as the solid black line.

Inspection of Fig.~\ref{fig:model_evolution} allows one to assess the relative
importance of the model refinements pursued in this article. Below roughly
\SI{30}{\MeV}, the original \marleyOld\ model provides a good approximation of
the full \marleyNew\ total cross section. At neutrino energies in the high tens
of \si{\MeV}, however, three new features of the \marleyNew\ model are seen to
have a substantial effect on $\sigma(E_\nu)$: the $\mathcal{O}(1/m_N)$ terms in
the tensor contraction (dashed red), the nuclear form factor (densely-dotted
yellow), and the forbidden multipoles (solid black). The first two of these
reduce the cross section while the forbidden multipoles provide an enhancement.
Although it lacks any forbidden contributions, the \marleyOld\ model
overestimates the allowed strength at high energies severely enough that the
full \marleyNew\ $\sigma(E_\nu)$ calculation only exceeds it just below
\SI{100}{\MeV}.

\subsection{Electron spectrum in the continuum}

\begin{figure}
\includegraphics[width=0.5\textwidth]{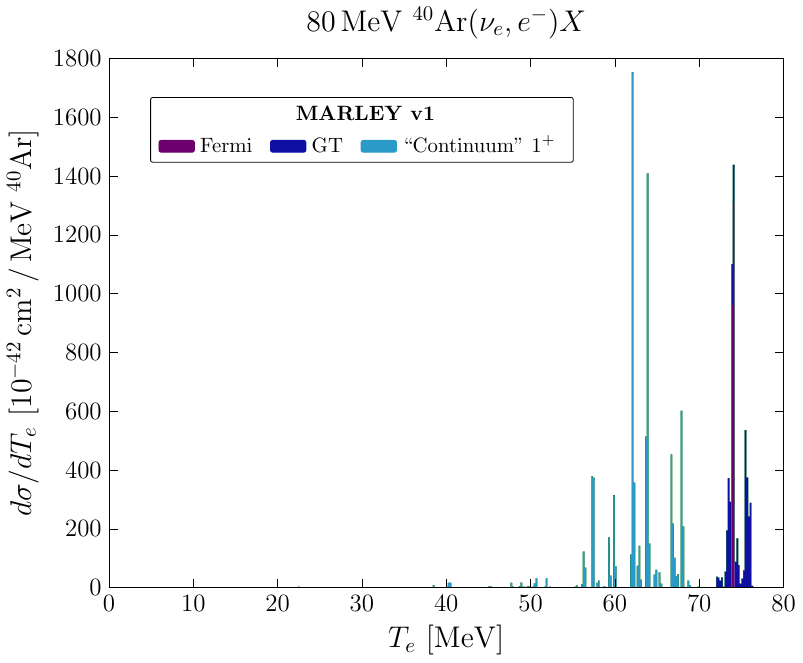}
\caption{\marley~\oldVersion\ differential cross section in electron kinetic
  energy for an \SI{80}{\MeV} incident neutrino.}
\label{fig:old_continuum}
\end{figure}

\begin{figure}
\includegraphics[width=0.5\textwidth]{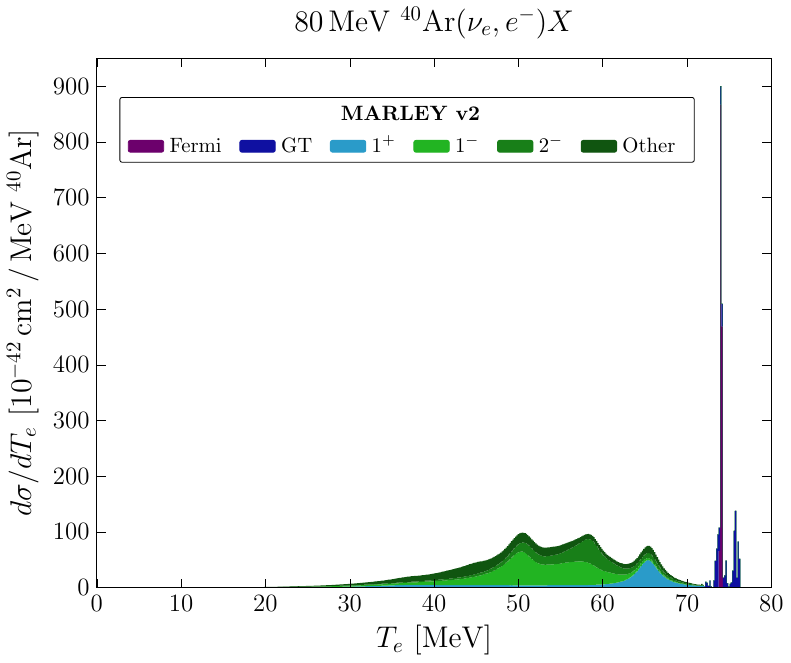}
\caption{\marley~\version\ differential cross section in electron kinetic energy
  for an \SI{80}{\MeV} incident neutrino.}
\label{fig:new_continuum}
\end{figure}

At high excitation energies, the \marleyNew\ model employs continuous nuclear
response functions computed with an HF-CRPA approach. This is a significant
change from the discrete level treatment used at all excitation energies in
\marleyOld, but its effect on the total cross section $\sigma(E_\nu)$ is
relatively modest (compare the densely-dashed yellow and dotted orange curves
in Fig.~\ref{fig:model_evolution}). The modeling differences become immediately
apparent, however, if one compares the predictions of the two code versions for
a monoenergetic neutrino source. Figure~\ref{fig:old_continuum} shows the
\marleyOld\ differential cross section as a function of outgoing electron
kinetic energy $T_e$ for an \SI{80}{\MeV} incident neutrino. Sharp peaks
corresponding to discrete nuclear energy levels are seen throughout the entire
distribution, and distinct colors are used to indicate the contribution of
different kinds of nuclear transitions. Violet is used to indicate the Fermi
transition, while both dark blue and cyan are used for GT transitions. The
separation between these two latter colors is made based on the decay modes
available for the $\isotope[40]{K}^*$ energy level populated by the primary
neutrino interaction. If the level is bound and must therefore de-excite via
$\gamma$-ray emission, then dark blue is used. If the level is unbound to
emission of one or more light nuclear fragments ($A \leq 4$), then cyan is
used. In the updated \marleyNew\ calculation, the discrete transitions
corresponding to the cyan portion of Fig.~\ref{fig:old_continuum} are
completely removed and replaced by the $1^+$ component of the continuum.

In Fig.~\ref{fig:new_continuum}, the complete \marleyNew\ prediction for the
same differential cross section is shown using a similar format. Although
discrete transitions are now modeled using the updated treatment from
Sec.~\ref{sec:discrete_xsec} rather than the AA, we nevertheless retain the
AA-inspired \textit{Fermi} (violet) and \textit{GT} (dark blue) labels to refer
to portions of the cross section in the discrete region arising from the first
and second terms of Eq.~\ref{eq:walecka_style_contraction}, respectively. These
labels will be used with the same meaning throughout the remainder of this
article.

The remaining colors in Fig.~\ref{fig:new_continuum} represent the portion of
the cross section modeled via the HF-CRPA approach, with the most prominent
$1^+$ (cyan), $1^-$ (light green), and $2^-$ (medium green) multipole
components shown separately from the sum of the rest (dark green). Although the
presence of a true continuum in the \marleyNew\ prediction is a striking
difference from the \marleyOld\ one, the cyan $1^+$ distribution in
Fig.~\ref{fig:new_continuum} is nevertheless concentrated in a similar
excitation energy region as the bulk of the cyan lines in
Fig.~\ref{fig:old_continuum}. The new calculation predicts substantially more
strength at moderate $T_e$ ($\sim 30$ to $\sim\SI{55}{\MeV}$) due to the
presence of forbidden multipoles in the continuum (green histograms), which are
entirely neglected in the \marleyOld\ model.

\subsection{Exclusive total cross sections}

\begin{figure}
\includegraphics[width=0.5\textwidth]{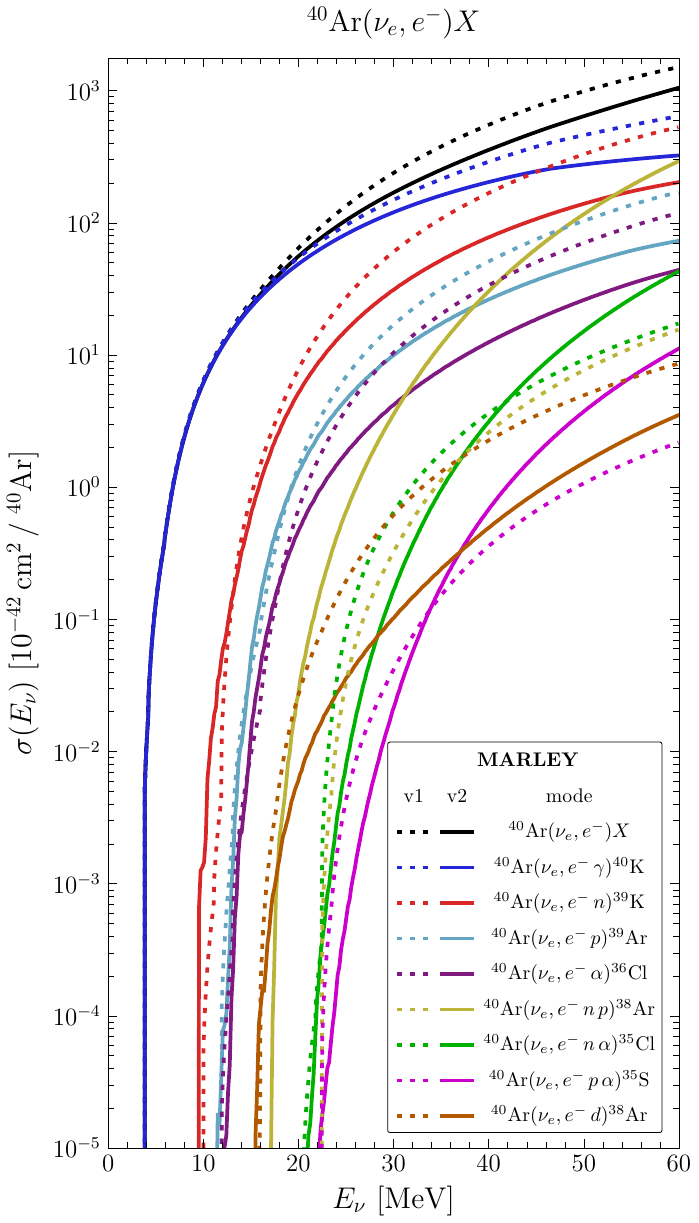}
\caption{Comparison of total cross sections predicted by
\marley~\oldVersion\ (\marleyOld, dashed lines) and
\marley~\version\ (\marleyNew, solid lines) as a function of neutrino
energy $E_\nu$. Predictions for both the inclusive total
cross section (black) and partial contributions from several specific
final states (other colors) are shown.}
\label{fig:excl_xsec}
\end{figure}

As a first examination of how the model updates presented above have changed
\marley's predictions for final states involving nuclear de-excitation
products, Fig.~\ref{fig:excl_xsec} compares several calculations of total cross
sections performed with the \marleyOld\ and \marleyNew\ treatments. A dashed
line style is used for all of the \marleyOld\ predictions, which are taken
directly from Ref.~\cite{marleyPRC}. Solid lines are used for all of the
\marleyNew\ results. The two black curves show predictions of the inclusive
total cross section for CC absorption of $\nu_e$ on \isotope[40]{Ar}. Both of
these were already plotted in Fig.~\ref{fig:model_evolution} using a different
format. The remaining colors are used to show contributions to the total cross
section from the most common nuclear de-excitation modes. The cross section for
final states in which only $\gamma$-rays are emitted from the outgoing nucleus
(referred to as $N\gamma$ below) is shown in blue. The remaining channels all
include any number of emitted $\gamma$-rays and are distinguished by their
hadronic content. These include cross sections for production of a single
neutron (red), proton (cyan), deuteron (dark orange), and $\alpha$ particle
(violet). Cross sections are also shown for production of one neutron and one
proton (yellow), one neutron and one $\alpha$ particle ($1n1\alpha$, green),
and one proton and one $\alpha$ particle ($1p1\alpha$, magenta).

Although the predictions of both \marley\ versions are very similar near the CC
reaction threshold, many of the plotted \marleyNew\ predictions are
considerably lower than \marleyOld\ at high neutrino energy. For the
$N\gamma$ channel as well as the channels involving emission of a
single neutron, proton, or $\alpha$ particle, this reduction arises primarily
due to the abandonment of the AA, under which the \marleyOld\ model neglected
the momentum transfer dependence of the nuclear responses. Incorporating this
dependence into the \marleyNew\ model leads to a suppression of the allowed
transitions as the neutrino energy rises and larger $\kappa$ values become
accessible. Although the continuum forbidden transitions considered in
\marleyNew\ partially restore some of the lost strength in the inclusive cross
section, their contribution is concentrated at relatively high excitation
energies where multi-fragment final states become possible. The increase from
\marleyOld\ to \marleyNew\ in the $1n1\alpha$ and $1p1\alpha$ cross sections
near \SI{60}{\MeV} can be attributed to the forbidden transitions, especially
the $1^{-}$ multipole.

While the corrections for non-zero $\kappa$ also play a role in reducing the
single-deuteron production cross section relative to \marleyOld, the most
important effect for this channel is a de-excitation model update. As discussed
in Sec.~\ref{sec:deex}, \marley~\oldVersion\ uses the highest tabulated nuclear
energy level as the continuum boundary when simulating de-excitations. In
contrast, \marley~\version\ ignores all tabulated levels above an excitation
energy $\ExContinuumThresh$ defined in Eq.~\ref{eq:continuum_threshold}. This
adjustment moves the lower bound of the continuum in $\isotope[39]{K}^*$ from
\SI{18.61}{\MeV} to \SI{6.38}{\MeV}, which leads to a corresponding drop of
roughly \SI{5}{\MeV} in the neutrino energy threshold for the $1n1p$ reaction.
Multi-step $1n1p$ emission quickly becomes dominant over single-step $d$
emission as the neutrino energy increases, further depleting the latter channel
beyond what might be expected from $\kappa$ corrections alone. With both a
lower threshold and a boost from continuum forbidden transitions in \marleyNew,
the $1n1p$ channel is markedly more prominent in the updated model. At $E_\nu =
\SI{60}{\MeV}$, it nearly overtakes the $N\gamma$ channel as the
leading contribution to the CC cross section.

While the results in Fig.~\ref{fig:excl_xsec} could have been obtained in
principle by generating a very large sample of \marley\ events with a uniform
neutrino energy spectrum, we adopted a more computationally efficient strategy
instead. For the portion of the cross section leading to discrete nuclear
levels, Eq.~\ref{eq:diff_xsec_discrete} was evaluated directly for each
kinematically accessible nuclear transition using the results from
Eqs.~\ref{eq:walecka_style_contraction} and~\ref{eq:B_with_F}. Integration over
$\cos\theta_\ell$ was handled numerically using the \marley\ implementation of
Clenshaw–Curtis quadrature~\cite{clenshaw1960}, and the partial cross sections
for each nuclear level were summed to obtain the total. In the \marleyNew\
model of the $\isotope[40]{Ar}(\nu_e, e^{-})\isotope[40]{K}^{*}$ reaction, only
bound nuclear levels are populated via discrete transitions. As a result, the
total cross section obtained via this procedure only contributes to the
inclusive (black) and $N\gamma$ (blue) curves in
Fig.~\ref{fig:excl_xsec}.

To evaluate exclusive predictions for the continuum portion of the cross
section, we created a dedicated program in which the \marley\ nuclear
de-excitation simulation can be run separately from the primary neutrino
interaction. A large sample of complete $\isotope[40]{K}^*$ de-excitation
events (including all binary decay steps) was generated using uniform
distributions for the initial nuclear spin $0 \leq J \leq 5$, parity $\Pi = \pm
1$, and excitation energy $E_x$ between the threshold for the continuum
$\ExContinuumThresh$ and \SI{100}{\MeV}. For each initial spin-parity $J^\Pi$,
a histogram of the total event counts as a function of $E_x$ was prepared using
a uniform bin width of \SI{0.1}{\MeV}. Similar histograms were also prepared
for the subset of simulated de-excitation events corresponding to each
final-state channel of interest. For an initial excitation energy $E_x$ lying
within the \BinIdx-th bin,
\begin{equation}
E_x^\BinIdx \leq E_x < E_x^{\BinIdx + 1} \,,
\end{equation}
the branching ratio $\BR$ for the final state
\FSIdx\ was estimated from the Monte Carlo simulation results via the expression
\begin{equation}
\BR(E_x, \FSIdx) \approx \frac{ n(E_x^\BinIdx, J, \Pi, \FSIdx) }
{ n(E_x^\BinIdx, J, \Pi) } \,,
\end{equation}
where $n(E_x^\BinIdx, J, \Pi, \FSIdx)$ is the number of simulated events
in the sample that matched the final-state definition \FSIdx\ and
had initial nuclear spin $J$, parity $\Pi$, and an excitation energy
$E_x$ within the
$i$-th bin. The denominator contains the total number of simulated
events with the same initial conditions, so that
\begin{equation}
n(E_x^\BinIdx, J, \Pi) = \sum_\FSIdx n(E_x^\BinIdx, J, \Pi, \FSIdx)
\end{equation}
when the sum includes all possible final states \FSIdx.

The total cross section for an exclusive final state \FSIdx\ was then evaluated
via the expression
\begin{equation}
  \sigma(E_\nu, \FSIdx) = \int \BR\big( E_x(\omega, \kappa), \FSIdx \big)
  \cdot \frac{ d^2 \sigma }{ d\omega \, d\kappa } \, d\omega \, d\kappa \,,
\end{equation}
where the initial excitation energy may be written in terms of the variables of
integration as
\begin{equation}
\label{eq:Ex_for_integrand}
  E_x( \omega, \kappa ) = \max\big( \ExContinuumThresh,
    \sqrt{ ( m_i + \omega )^2 - \kappa^2 } - m_f^\mathrm{g.s.} \big) \,.
\end{equation}
Taking the maximum of the calculated excitation energy and \ExContinuumThresh\
in Eq.~\ref{eq:Ex_for_integrand} incorporates the correction discussed in
Sec.~\ref{sec:sub_continuum} above. Integration over $\omega$ and $\kappa$ was
performed using precomputed tables of the nuclear response functions and the
two-dimensional trapezoidal rule. The differential cross section in these
variables was obtained using the expression in
Eq.~\ref{eq:diff_xsec_continuum_DeltaCorrected} and the Jacobian transformation
\begin{equation}
  \frac{ d^2 \sigma }{ d\omega \, d\kappa } =
  \frac{ \kappa }{ |\mathbf{\pnu}| \, |\mathbf{\plep}| }
  \cdot \frac{ d^2\sigma }{ dT_\ell \, d\cos\theta_\ell } \,.
\end{equation}

\subsection{Multipole inclusive cross sections}

\begin{figure}
\includegraphics[width=0.5\textwidth]{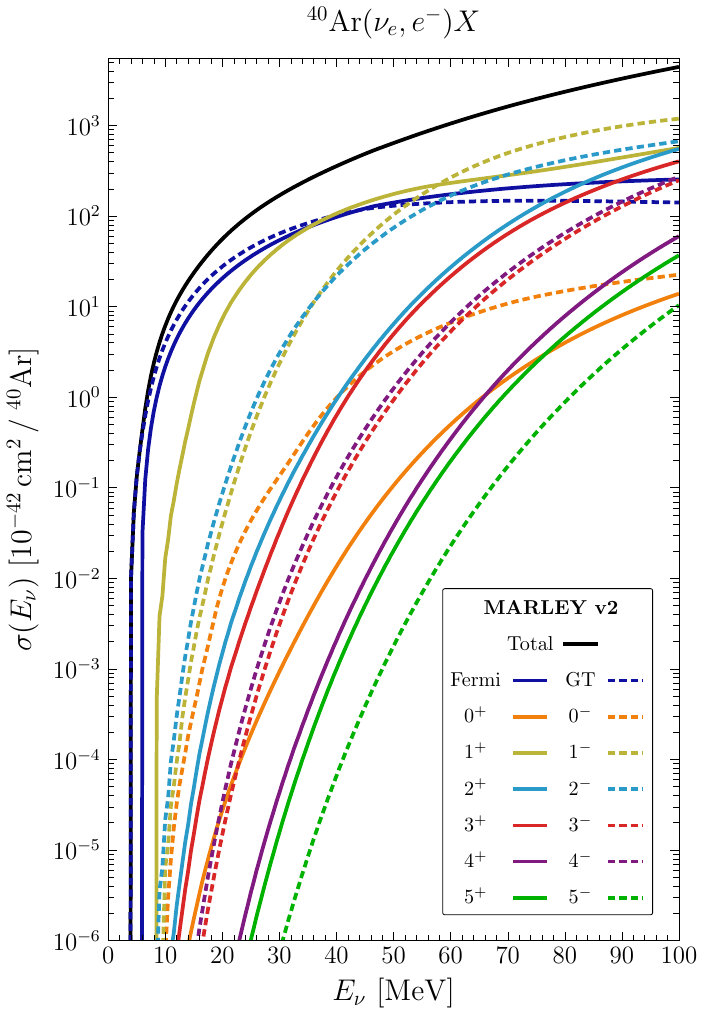}
\caption{Discrete and continuum multipole contributions to the
inclusive total cross section.}
\label{fig:tot_xsec_multipole}
\end{figure}

Figure~\ref{fig:tot_xsec_multipole} once again plots \marley\ total cross
sections as a function of neutrino energy. However, in contrast to
Fig.~\ref{fig:excl_xsec}, only \marleyNew\ calculations are shown here, and the
total CC inclusive cross section (solid black) is decomposed into multipole
contributions rather than into specific final states. For the $0^+$ and $1^+$
multipoles, our model considers both discrete (Sec.~\ref{sec:discrete_xsec})
and continuum (Sec.~\ref{sec:continuum_xsec}) nuclear transitions. In
Fig.~\ref{fig:tot_xsec_multipole}, the Fermi and GT components of the cross
section for discrete transitions are plotted as the solid and dashed blue
curves, respectively. The remaining curves show the continuum cross section
components, which bear numerical labels giving the multipole order and parity.
For all continuum multipoles plotted, color is used to represent the multipole
order, while a solid (dashed) line style indicates positive (negative) parity.

Below roughly \SI{30}{\MeV}, the cross section is strongly dominated by allowed
transitions (GT, Fermi, and continuum $1^+$). Around \SI{50}{\MeV}, the
forbidden $1^-$ and $2^-$ continuum multipoles both begin to become comparable
to the allowed components, with $1^-$ growing to be the largest single
component of the cross section in the high tens of \si{\MeV}. At
\SI{100}{\MeV}, several forbidden multipoles make sizeable contributions.

\subsection{Excitation energy in the continuum}

\begin{figure}
\includegraphics[width=0.48\textwidth]{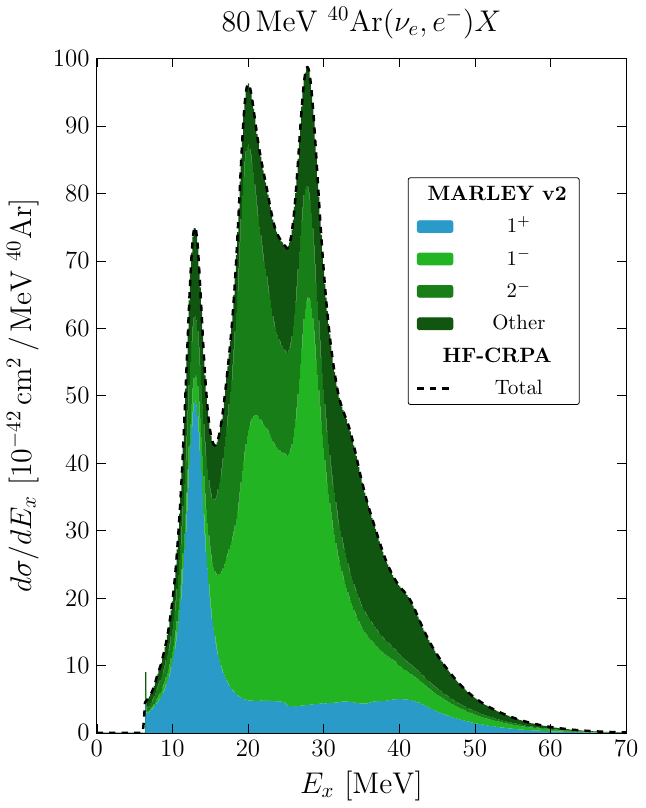}
  \caption{Continuum-only CC differential cross section as a function of
  $\isotope[40]{K}^*$ excitation energy $E_x$ for an \SI{80}{\MeV} incident
  $\nu_e$. The HF-CRPA and \marley\ calculations agree closely except where the
  latter reassigns sub-continuum strength to the threshold
  $\ExContinuumThresh$.}
\label{fig:diff_xsec_Ex_continuum}
\end{figure}

Figure~\ref{fig:diff_xsec_Ex_continuum} shows the continuum portion of the
\marleyNew\ differential cross section as a function of nuclear excitation
energy $E_x$ above the $\isotope[40]{K}$ ground state. The stacked histogram is
shaded to indicate contributions from the dominant $1^+$ (cyan), $1^-$ (light
green), and $2^-$ (medium green) multipoles, with the remainder combined into
the dark green region. The black dashed line shows an analytic calculation of
the total cross section using the HF-CRPA nuclear response functions. From the
expression in Eq.~\ref{eq:diff_xsec_continuum_DeltaCorrected}, one may use the
Jacobian transformation
\begin{equation}
\frac{ d^2\sigma }{ dE_x \, d\cos\theta_\ell }
  = \frac{ m_f }{ E_\mathrm{tot} - \beta_\ell^{-1}
  \, |\mathbf{\pnu}| \, \cos\theta_\ell }
  \cdot \frac{ d^2\sigma }{ dT_\ell \, d\cos\theta_\ell } \,,
\end{equation}
and integrate over scattering angle to obtain $d\sigma/dE_x$. In the analytic
calculation, Lorentzian broadening (Sec.~\ref{sec:lorentzian_folding}) and the
energy shift $\DeltaCRPA$ (Sec.~\ref{sec:crpa_energy_balance}) have both been
applied, but the reassignment of sub-continuum strength
(Sec.~\ref{sec:sub_continuum}) has been left out. The impact of this
reassignment can be seen in the \marley\ histogram's single-bin spike that
occurs at the continuum threshold $E_x = \ExContinuumThresh =
\SI{6.438}{\MeV}$. Although a noticeable (and intentional) discrepancy exists
between the \marley\ and HF-CRPA results in this bin, the two results are
otherwise in excellent agreement.

\subsection{Supernova neutrinos}

\begin{figure}
\includegraphics[width=0.5\textwidth]{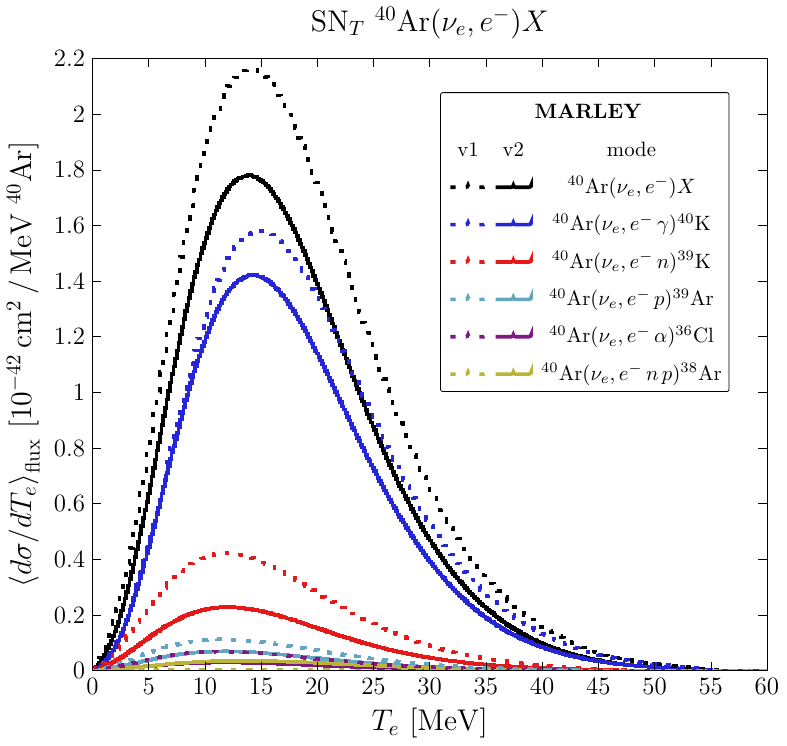}
\caption{Comparison of \marleyOld\ and \marleyNew\ differential cross sections
  as for supernova $\nu_e$ a function of the electron kinetic energy $T_e$. The
  results are flux-averaged using the time-integrated SN$_T$ $\nu_e$ energy
  spectrum described in the text.}
\label{fig:SN_Te_xsec}
\end{figure}

\begin{figure}
\includegraphics[width=0.5\textwidth]{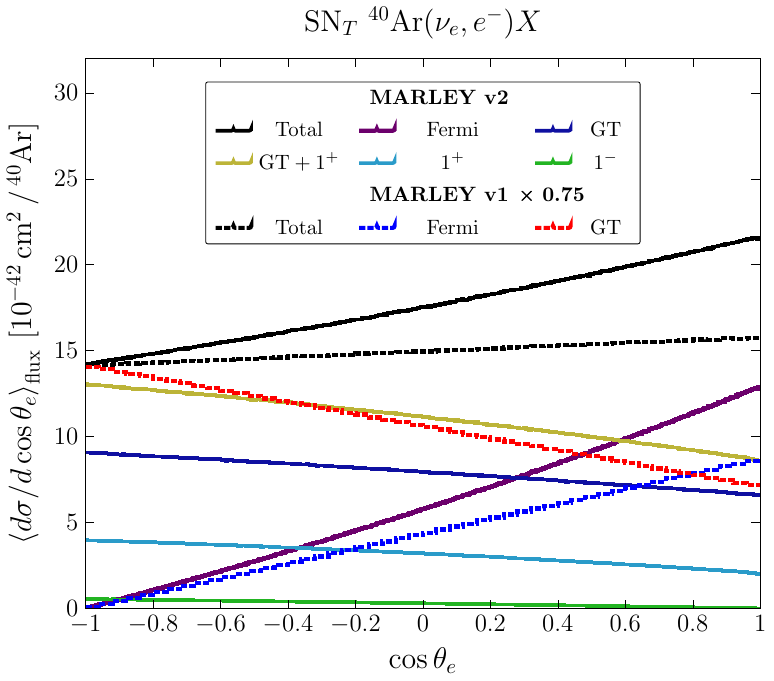}
\caption{Flux-averaged differential cross sections
  for supernova $\nu_e$ as a function of electron scattering cosine.}
\label{fig:SN_cos_xsec}
\end{figure}

\begin{figure}
\includegraphics[width=0.5\textwidth]{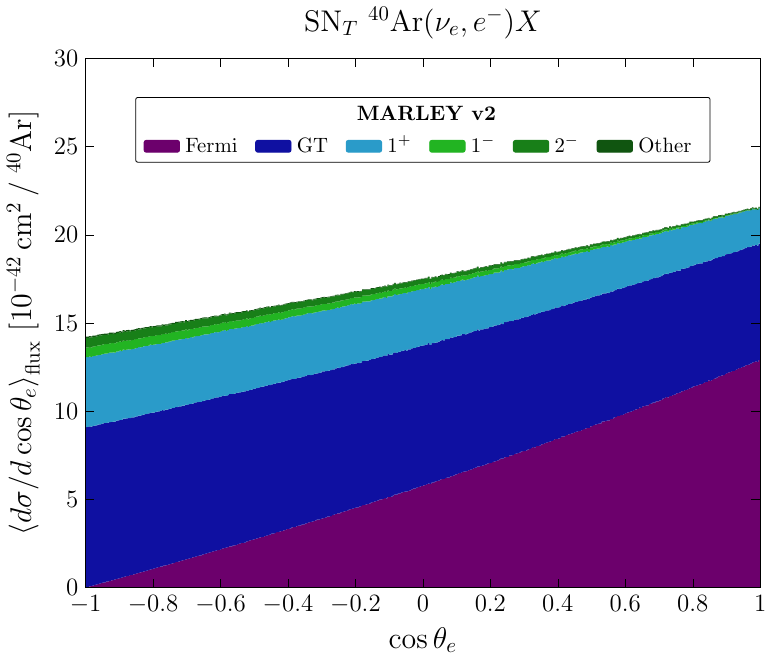}
\caption{Stacked contributions to the flux-averaged inclusive \marleyNew\
  differential cross section for supernova $\nu_e$ as a function of electron
  scattering cosine.}
\label{fig:SN_stacked_cos_xsec}
\end{figure}

To examine the impact of our model refinements on predictions for supernova
neutrino interactions, we adopt the time-integrated $\nu_e$
energy spectrum from Ref.~\cite{Nikrant2018}, which has the functional form
\newcommand{\Eavg}{ \left<E_\nu\right> }
\begin{equation}
  \label{eq:SN_flux}
  \phi_{\mathrm{SN}_T}(E_\nu)
  \propto \left( \frac{ E_\nu }{ \Eavg } \right)^{\alpha}
  \exp\left[ -(\alpha + 1) \, \frac{ E_\nu }{ \Eavg } \right] \,.
\end{equation}
Here $\Eavg = \SI{14.1}{\MeV}$ is the mean $\nu_e$ energy and $\alpha = 2.67$
is a shape parameter. We refer to results obtained using this energy spectrum
with the label SN$_T$. Although there is significant variation between
calculations of supernova spectra in the literature, we choose this particular
shape as a representative example and for consistency with a prior \marley\
study~\cite{marleyPRC}.

In Fig.~\ref{fig:SN_Te_xsec}, the SN$_T$ supernova $\nu_e$ energy spectrum from
Eq.~\ref{eq:SN_flux} is used to compute flux-averaged differential cross
sections as a function of outgoing electron kinetic energy $T_e$. Both
inclusive and exclusive predictions from \marleyOld\ and \marleyNew\ are
plotted using the same line styles and color scheme as in
Fig.~\ref{fig:excl_xsec}. In addition to an overall reduction of the inclusive
cross section by about 20\%, there is also a slight change of shape that
reduces the mean $T_e$ by \SI{0.4}{\MeV}. With the exception of the $1n1p$
distribution, which has grown to 2\% of the total in \marleyNew, all exclusive
$T_e$ cross sections shown are reduced in the switch from \marleyOld\ to
\marleyNew. While the single-fragment emission channels all make smaller
fractional contributions (e.g., $1n$ is 18\% of the \marleyOld\ total versus
12\% of \marleyNew), the $N\gamma$ final state becomes slightly more dominant
(73\% of the \marleyOld\ total versus 80\% of \marleyNew). Below $T_e =
\SI{8}{\MeV}$, the updated $N\gamma$ prediction is almost unchanged from
\marleyOld. This region of the blue distribution is dominated by the low-energy
portion of the $\nu_e$ spectrum, where the corrections for $\kappa$ dependence
introduced in \marleyNew\ have the smallest impact.

Figure~\ref{fig:SN_cos_xsec} compares the \marleyOld\ and \marleyNew\ angular
distributions averaged over the SN$_T$ flux. To facilitate shape comparisons,
the \marley~\oldVersion\ predictions (dashed lines) for the total cross section
(black) and the Fermi (blue) and GT components (red) are all multiplied by a
scaling factor of $0.75$. Solid lines are used to draw the \marley~\version\
predictions, including the total (black) and most important partial
contributions. The Fermi component of the \marleyNew\ differential cross
section (violet) is directly comparable to its \marleyOld\ version. High-lying
GT transitions in \marleyOld, on the other hand, have been replaced with the
$1^+$ multipole of the HF-CRPA treatment in \marleyNew. The appropriate updated
angular distribution for a direct comparison to the \marleyOld\ GT prediction
is therefore the sum of the \marleyNew\ cross sections arising from continuum
$1^+$ transitions (cyan) and discrete GT transitions (dark blue). This sum is
plotted as the yellow line. Despite major differences in the underlying
calculations, the discrete GT and continuum $1^+$ results nevertheless have
similar shapes. As an example of the angular distribution for the forbidden
transitions, the continuum $1^{-}$ multipole component of the cross section is
plotted in green.

The suppression of the allowed portion of the SN$_T$ cross section in
\marleyNew\ is stronger at backward angles and is not fully compensated by the
addition of forbidden transitions. As a result, the nearly-uniform angular
distribution predicted by the AA now noticeably favors the forward direction
after our model updates. Although recent DUNE studies of supernova pointing
deliberately ignored the nuclear CC reaction in favor of an analysis using
neutrino-electron elastic scattering event candidates~\cite{DUNE:2024ptd}, the
new \marley\ prediction for the CC angular distribution may motivate a
different strategy in future work on the subject. Given the especially strong
preference for forward scattering angles in the Fermi distribution, selection
criteria that attempt to identify CC events generated by this transition (via
$\gamma$-ray \textit{blip} reconstruction~\cite{MicroBooNE:2024prh}) may prove
useful for maximizing DUNE's supernova pointing capability. For applications in
which neutrino calorimetry is important, the known \SI{4.384}{\MeV} excitation
energy of the isobaric analog state accessed in the Fermi transition will also
enable greatly improved energy resolution for this subsample of events.

As an alternative visualization of the relative importance of the \marleyNew\
cross section components in shaping the electron angular distribution,
Fig.~\ref{fig:SN_stacked_cos_xsec} shows the overall differential cross section
in $\cos\theta_e$ as a stacked histogram. Separate colors are used to represent
the contributions from the Fermi (violet), GT (dark blue), and continuum $1^+$
(cyan), $1^-$ (light green), and $2^-$ (medium green) components. A very small
portion of the cross section due to other continuum multipoles (dark green) is
also included in the figure. Although corrections beyond the AA have an
important impact on the overall cross-section scale and details of the lepton
kinematic distributions, it is still immediately apparent from
Fig.~\ref{fig:SN_stacked_cos_xsec} that allowed transitions (violet, dark blue,
and cyan) remain overwhelmingly dominant in \marleyNew\ predictions for the
SN$_T$ neutrino spectrum.

\subsection{Muon decay-at-rest neutrinos}

\begin{figure}
\includegraphics[width=0.5\textwidth]{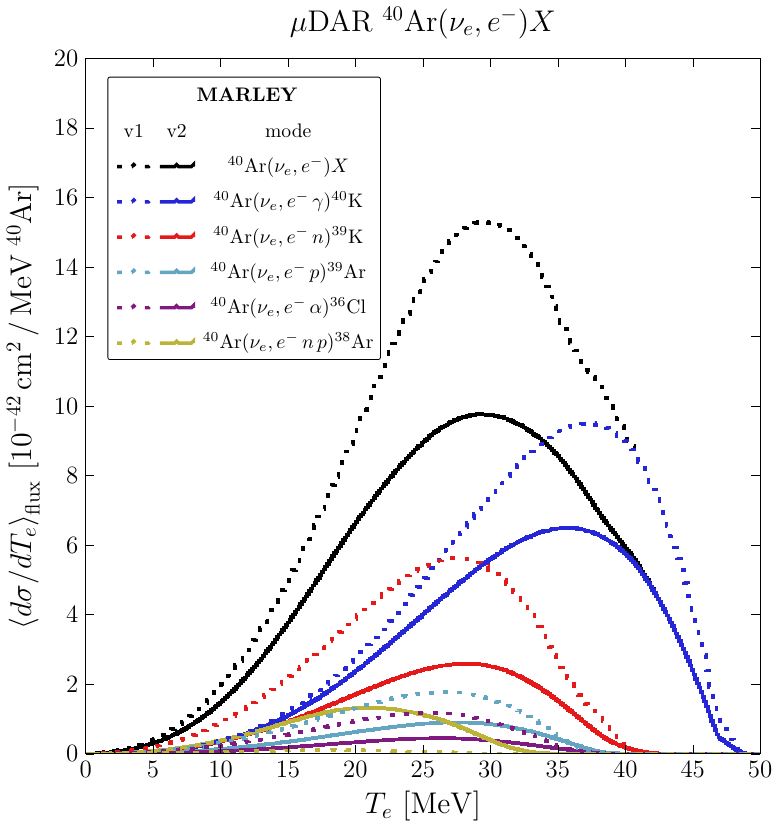}
\caption{Flux-averaged differential cross section in electron kinetic energy.}
\label{fig:muDAR_Te_xsec}
\end{figure}

\begin{figure}
\includegraphics[width=0.5\textwidth]{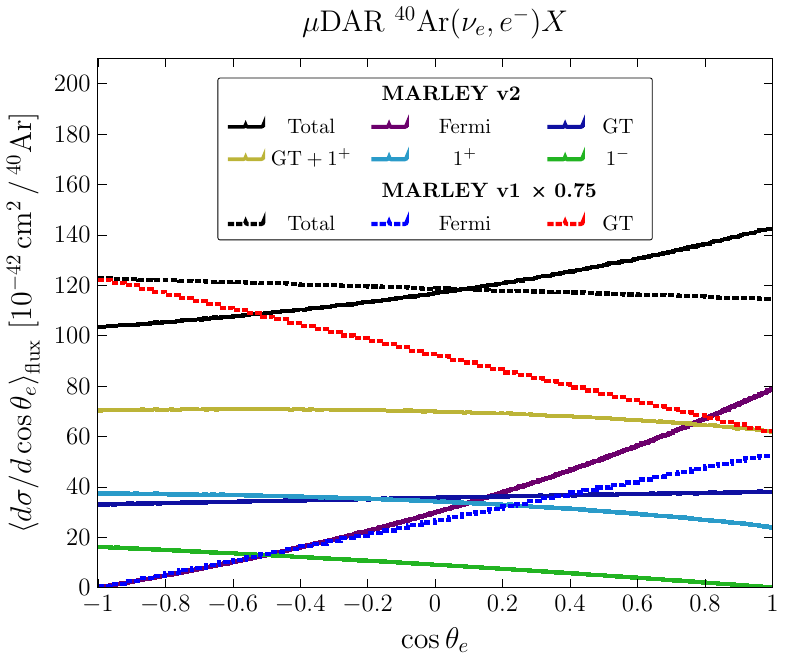}
\caption{Flux-averaged differential cross sections in electron scattering
angle.}
\label{fig:muDAR_cos_xsec}
\end{figure}

\begin{figure}
\includegraphics[width=0.5\textwidth]{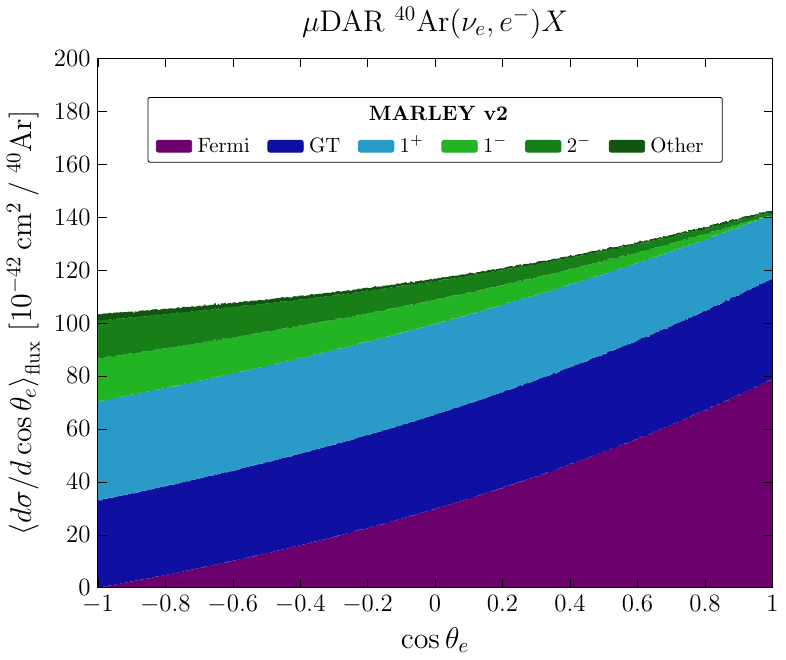}
\caption{Stacked contributions to total inclusive flux-averaged differential
cross section in electron scattering angle.}
\label{fig:muDAR_stacked_cos_xsec}
\end{figure}

To measure neutrino cross sections at the tens-of-\si{\MeV} energies relevant
for supernovae, the experimental community has typically used accelerator
facilities that produce large numbers of stopped antimuons. When an antimuon
at rest undergoes the decay
\begin{equation}
\mu^{+} \rightarrow e^{+} + \nu_e + \bar{\nu}_\mu\,,
\end{equation}
the emitted $\nu_e$ has the energy distribution~\cite{Kosmas2017}
\begin{equation}
\label{eq:mudar}
\phi_{\mu\mathrm{DAR}}(E_\nu)
  \propto E_\nu^2 \, m_\mu^{-4} \, (m_\mu - 2\,E_\nu) \,,
\end{equation}
where $m_\mu$ is the muon mass and
\begin{equation}
0 < E_\nu < m_\mu/2 \,.
\end{equation}
Although CC cross-section data for an argon target has not been obtained yet
using muon decay-at-rest ($\mu$DAR) electron neutrinos, the COHERENT
collaboration has recently performed direct tests of \marley\ predictions for
lead~\cite{ninspb} and iodine~\cite{coherent:2023ffx} using $\mu$DAR neutrinos
from the Spallation Neutron Source (SNS) at Oak Ridge National Laboratory.
Additional COHERENT measurements of low-energy neutrino cross sections at the
SNS, including the CC reaction on argon, are expected in the coming
years~\cite{COHERENT:2026ewu}. Other experimental collaborations, including
CCM~\cite{CCM:2021leg} and SBND~\cite{sbndposter}, have also expressed interest
in performing a cross-section measurement using $\mu$DAR $\nu_e$ CC
interactions on argon.

In anticipation of direct experimental measurements, we present predictions for
CC $\nu_e$-\isotope[40]{Ar} differential cross sections in
Figs.~\ref{fig:muDAR_Te_xsec}--\ref{fig:muDAR_stacked_cos_xsec} that are
calculated with \marley~\version\ and averaged over the $\mu$DAR energy
spectrum from Eq.~\ref{eq:mudar}. Figure~\ref{fig:muDAR_Te_xsec} compares the
\marleyOld\ and \marleyNew\ differential cross sections as a function of the
electron kinetic energy $T_e$. The plot format is identical to
Fig.~\ref{fig:SN_Te_xsec}. The same general features from the supernova $\nu_e$
results are also seen here, although the overall suppression of the total cross
section (now 33\% lower than in \marleyOld) is more dramatic due to the higher
mean neutrino energy. Fragment emission, particularly in the $1n$ and $1n1p$
channels, is considerably more important for the $\mu$DAR spectrum than for
SN$_T$. Although the $1n$ channel's share of the total cross section drops from
29\% in \marleyOld\ to 20\% in \marleyNew, the $1n1p$ portion rises from 0.4\%
to 8.8\%, leaving the fraction of events involving emission of at least one
neutron (30.4\% in \marleyOld\ versus 30.7\% in \marleyNew) nearly identical.

Figure~\ref{fig:muDAR_cos_xsec} shows $\mu$DAR differential cross sections as a
function of electron scattering cosine using the same format as
Fig.~\ref{fig:SN_cos_xsec}. Here the \marleyNew\ total cross section (solid
black) is roughly 40\% larger at $\cos\theta_e = 1$ than at $\cos\theta_e =
-1$, in contrast to the \marleyOld\ prediction (dashed black) that is slightly
higher at backward angles. As in the SN$_T$ angular distributions, the
corrections for non-zero momentum transfer added in \marleyNew\ suppress the
allowed cross sections more strongly in the backward direction, thus allowing
the forward shape of the Fermi cross section (violet) to dominate even in the
presence of forbidden transitions.

The enhanced role of the $1^-$ and $2^-$ forbidden multipoles for $\mu$DAR
$\nu_e$ interactions versus SN$_T$ $\nu_e$ may be seen by comparing the stacked
histogram in Fig.~\ref{fig:muDAR_stacked_cos_xsec} with the one from
Fig.~\ref{fig:SN_stacked_cos_xsec}. Although the contribution of forbidden
transitions (green histograms) remains sub-leading for the $\mu$DAR spectrum,
it has grown to 14.5\% of the total from a mere 3.4\% for SN$_T$.

\subsection{Flux-averaged total cross sections}

\pgfplotstableread{tables/dar-modes.txt}{\mudarTable}
\pgfplotstableread{tables/nikrant_sn-modes.txt}{\snTable}
\pgfplotstableread{tables/mono-80MeV-modes.txt}{\monoTable}

\newcommand{\myNum}[1]{\num[ round-mode = places, round-precision = 1,
  retain-zero-exponent = true, scientific-notation = true ]{ #1 }}%
\newcommand{\GetTableValue}[3]{%
  \pgfplotstablegetelem{#1}{#2}\of{#3}\myNum{\pgfplotsretval}%
}%
\newcommand{\GetTableValueWithUnc}[3]{%
  \pgfplotstablegetelem{#1}{#2}\of{#3}\edef\myVal{\pgfplotsretval}%
  \pgfplotstablegetelem{#1}{#2_MCstatUnc}\of{#3}\edef\myErr{\pgfplotsretval}%
  \myNum{ \myVal   \pm   \myErr }%
}%
%
\newcommand{\ReactionRow}[2]{%
  {$#1$} &
  \GetTableValue{#2}{tot}{\snTable} &
  \GetTableValue{#2}{tot}{\mudarTable} &
  \GetTableValue{#2}{tot}{\monoTable} \\
}%
%
\newcommand{\PartialInclusiveRow}[2]{%
  {#1} &
  \GetTableValue{67}{#2}{\snTable} &
  \GetTableValue{67}{#2}{\mudarTable} &
  \GetTableValue{67}{#2}{\monoTable} \\
}%

\begin{table}
\caption{Flux-averaged total cross sections
  ($\SI{e-42}{\centi\meter\squared} / \isotope[40]{Ar}$) for several
  reference $\nu_e$ spectra.}
\label{tab:channel_xsecs}
\begin{tabular}{llll}
\doubletoprule{1}
Channel & {SN$_T$} & {$\mu$DAR} & {\SI{80}{\MeV}} \\
\midrule
\ReactionRow{\isotope[40]{Ar}(\nu_e, e^{-})X}{67}
\ReactionRow{\isotope[40]{Ar}(\nu_e, e^{-} \, \gamma)\isotope[40]{K}}{66}
\ReactionRow{\isotope[40]{Ar}(\nu_e, e^{-} \, n)\isotope[39]{K}}{7}
\ReactionRow{\isotope[40]{Ar}(\nu_e, e^{-} \, p)\isotope[39]{Ar}}{11}
\ReactionRow{\isotope[40]{Ar}(\nu_e, e^{-} \, d)\isotope[38]{Ar}}{3}
\ReactionRow{\isotope[40]{Ar}(\nu_e, e^{-} \, t)\isotope[37]{Ar}}{12}
\ReactionRow{\isotope[40]{Ar}(\nu_e, e^{-} \, h)\isotope[37]{Cl}}{4}
\ReactionRow{\isotope[40]{Ar}(\nu_e, e^{-} \, \alpha)\isotope[36]{Cl}}{1}
\ReactionRow{\isotope[40]{Ar}(\nu_e, e^{-} \, 2n)\isotope[38]{K}}{23}
\ReactionRow{\isotope[40]{Ar}(\nu_e, e^{-} \, n \, p)\isotope[38]{Ar}}{9}
\ReactionRow{\isotope[40]{Ar}(\nu_e, e^{-} \, n \, d)\isotope[37]{Cl}}{2}
\ReactionRow{\isotope[40]{Ar}(\nu_e, e^{-} \, n \, \alpha)\isotope[35]{Cl}}{5}
\ReactionRow{\isotope[40]{Ar}(\nu_e, e^{-} \, 2p)\isotope[38]{Cl}}{25}
\ReactionRow{\isotope[40]{Ar}(\nu_e, e^{-} \, p \, \alpha)\isotope[35]{S}}{8}
\ReactionRow{\isotope[40]{Ar}(\nu_e, e^{-} \, 2\alpha)\isotope[32]{Si}}{19}
\ReactionRow{\isotope[40]{Ar}(\nu_e, e^{-} \, 2n \, p)\isotope[37]{Ar}}{10}
\ReactionRow{\isotope[40]{Ar}(\nu_e, e^{-} \, n \, 2p)\isotope[37]{Cl}}{24}
\ReactionRow{\isotope[40]{Ar}(\nu_e, e^{-} \, X)\isotope[32]{P}}{32}
\ReactionRow{\isotope[40]{Ar}(\nu_e, e^{-} \, X)\isotope[34]{S}}{41}
\ReactionRow{\isotope[40]{Ar}(\nu_e, e^{-} \, X)\isotope[35]{Cl}}{43}
\ReactionRow{\isotope[40]{Ar}(\nu_e, e^{-} \, X)\isotope[35]{S}}{45}
\ReactionRow{\isotope[40]{Ar}(\nu_e, e^{-} \, X)\isotope[36]{Cl}}{47}
\ReactionRow{\isotope[40]{Ar}(\nu_e, e^{-} \, X)\isotope[37]{Ar}}{50}
\ReactionRow{\isotope[40]{Ar}(\nu_e, e^{-} \, X)\isotope[37]{Cl}}{51}
\ReactionRow{\isotope[40]{Ar}(\nu_e, e^{-} \, X)\isotope[38]{Ar}}{54}
\ReactionRow{\isotope[40]{Ar}(\nu_e, e^{-} \, X)\isotope[38]{Cl}}{55}
\ReactionRow{\isotope[40]{Ar}(\nu_e, e^{-} \, X)\isotope[38]{K}}{56}
\ReactionRow{\isotope[40]{Ar}(\nu_e, e^{-} \, 0n)X}{0}
\ReactionRow{\isotope[40]{Ar}(\nu_e, e^{-} \, 1n)X}{6}
\ReactionRow{\isotope[40]{Ar}(\nu_e, e^{-} \, 2n)X}{22}
\doublebottomrule{1}
\end{tabular}
\end{table}

\begin{table}
\caption{Multipole inclusive total cross sections
  ($\SI{e-42}{\centi\meter\squared} / \isotope[40]{Ar}$)
  averaged over several reference $\nu_e$ spectra.}
\label{tab:inclusive_multipole_xsecs}
\begin{tabular}{llll}
\doubletoprule{1}
Component & {SN$_T$} & {$\mu$DAR} & {\SI{80}{\MeV}} \\
\midrule
\ReactionRow{\isotope[40]{Ar}(\nu_e, e^{-})X}{67}
\PartialInclusiveRow{Fermi}{discrete-F}
\PartialInclusiveRow{Gamow-Teller}{discrete-GT}
\PartialInclusiveRow{Continuum $0^{+}$}{continuum-0pos}
\PartialInclusiveRow{Continuum $0^{-}$}{continuum-0neg}
\PartialInclusiveRow{Continuum $1^{+}$}{continuum-1pos}
\PartialInclusiveRow{Continuum $1^{-}$}{continuum-1neg}
\PartialInclusiveRow{Continuum $2^{+}$}{continuum-2pos}
\PartialInclusiveRow{Continuum $2^{-}$}{continuum-2neg}
\PartialInclusiveRow{Continuum $3^{+}$}{continuum-3pos}
\PartialInclusiveRow{Continuum $3^{-}$}{continuum-3neg}
\PartialInclusiveRow{Continuum $4^{+}$}{continuum-4pos}
\PartialInclusiveRow{Continuum $4^{-}$}{continuum-4neg}
\PartialInclusiveRow{Continuum $5^{+}$}{continuum-5pos}
\PartialInclusiveRow{Continuum $5^{-}$}{continuum-5neg}
\doublebottomrule{1}
\end{tabular}
\end{table}

As a quick reference for the relative importance of various channels in the
\marleyNew\ model of the $\isotope[40]{Ar}(\nu_e, e^{-})\isotope[40]{K}^*$
reaction, we provide two tables of flux-averaged total cross sections for the
SN$_T$ spectrum, the $\mu$DAR spectrum, and a monoenergetic \SI{80}{\MeV}
neutrino source. Table~\ref{tab:channel_xsecs} reports cross sections for
various exclusive and semi-inclusive final states. A similar table for the
\marleyOld\ model is available in Ref.~\cite{marleyPRC}.
Table~\ref{tab:inclusive_multipole_xsecs} shows a breakdown of the inclusive
total cross section into Fermi, GT, and continuum multipole components.

\subsection{Neutrino energy reconstruction}

\begin{figure}
\includegraphics[width=0.5\textwidth]{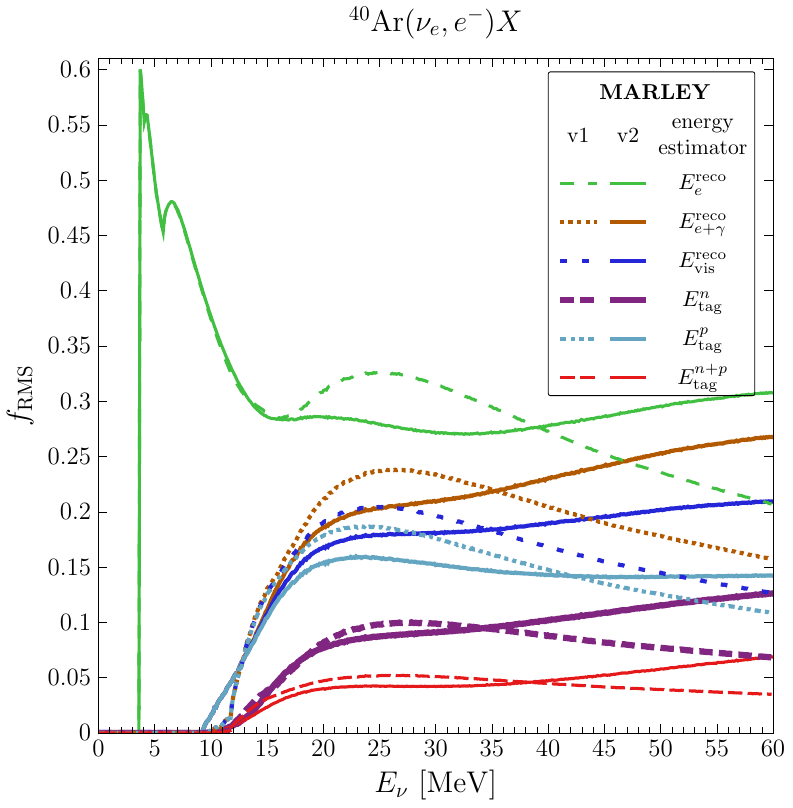}
\caption{Fractional RMS neutrino energy resolution.}
\label{fig:Ev_resolution}
\end{figure}

To examine the \marley~\oldVersion\ model's implications for neutrino energy
reconstruction, Ref.~\cite{marleyPRC} defined several energy estimators
representing different assumptions about future LArTPC capabilities. Idealized
reconstruction performance (accounting for physics-driven limitations but not
details of the detector response) was then assessed by calculating the
fractional root-mean-square (RMS) resolution on the neutrino energy. A similar
metric computed using a full detector simulation was used by the DUNE
collaboration for initial studies of sensitivity to a supernova neutrino
burst~\cite[Fig.~7]{Abi2020}.

In Fig.~\ref{fig:Ev_resolution}, we compare the fractional RMS resolution on
$E_\nu$ calculated with the new \marley~\version\ model to the results obtained
with the prior code version. Curves are shown for all six energy estimators
originally defined in Ref.~\cite{marleyPRC}. Three of these are constructed by
summing the nuclear mass difference $E_\mathrm{bind}^\mathrm{min}$ between the
\isotope[40]{K} and \isotope[40]{Ar} ground states,
\begin{align}
  \nonumber
  E_\mathrm{bind}^\mathrm{min}
  &\equiv \mNuclear(\isotope[40]{K}) - \mNuclear(\isotope[40]{Ar}) \\
  \nonumber
  &\approx \mAtom(\isotope[40]{K}) - \mAtom(\isotope[40]{Ar})
  - m_e \\
  &= \SI{0.993}{\MeV} \,,
\end{align}
with the energies of one or more final-state particles. The
$E_e^\mathrm{reco}$ estimator adds only the total energy of the
outgoing electron,
\begin{equation}
  E_e^\mathrm{reco} \equiv E_\mathrm{bind}^\mathrm{min}
    + E_e \,,
\end{equation}
the $E_{e+\gamma}^\mathrm{reco}$ estimator also includes
the energies of all final-state $\gamma$-rays,
\begin{equation}
E_{e+\gamma}^\mathrm{reco} \equiv E_e^\mathrm{reco}
  + \sum_\gamma E_\gamma \,,
\end{equation}
and the $E_\mathrm{vis}^\mathrm{reco}$ estimator includes the prior terms as
well as the kinetic energies of all outgoing charged hadrons:
\begin{equation}
  E_\mathrm{vis}^\mathrm{reco} \equiv E_{e+\gamma}^\mathrm{reco}
    + \sum_{h^\pm} T_{h^\pm} \,.
\end{equation}
The other energy estimators build upon $E_\mathrm{vis}^\mathrm{reco}$
by adding a correction for the binding energy of emitted nuclear fragments.
If we define the symbol $\delta_n$ ($\delta_{h^\pm}$) to be one for
an event containing a final-state neutron (proton or light ion), then
we may write the remaining energy estimators as
\begin{align}
  E_\mathrm{tag}^\mathrm{n} &\equiv E_\mathrm{vis}^\mathrm{reco}
    + \delta_n \, S_n \\[0.5\baselineskip]
  E_\mathrm{tag}^\mathrm{p} &\equiv E_\mathrm{vis}^\mathrm{reco}
    + \delta_{h^\pm} \, S_p \\[0.5\baselineskip]
  \nonumber
  E_\mathrm{tag}^{n+p} &\equiv E_\mathrm{vis}^\mathrm{reco}
    + \delta_n \, (1 - \delta_{h^\pm}) \,
      S_n
    \\
    & \;\;\;\;\;\; + \delta_{h^\pm} \, (1 - \delta_n) \,
      S_p
    + \delta_n \, \delta_{h^\pm} \,
      S_{np} \,.
\end{align}
Here the neutron (proton) separation energy $S_n = \SI{7.80}{\MeV}$
($S_p = \SI{7.58}{\MeV}$) is calculated as
described in Sec.~\ref{sec:sub_continuum}, and
\begin{align}
  \nonumber
  S_{np} &\equiv \mNuclear(\isotope[38]{Ar}) + m_n + m_p
  - \mNuclear(\isotope[40]{K}) \\
  \nonumber
  \approx \mAtom&(\isotope[38]{Ar}) + m_n + m_p + m_e
  - \mAtom(\isotope[40]{K}) \\
  &= \SI{14.18}{\MeV} \,,
\end{align}
where $m_n$ ($m_p$) is the neutron (proton) mass. More details about these
estimators and the calculation of the fractional RMS neutrino energy resolution
$f_\mathrm{RMS}$ from \marley\ Monte Carlo events are given in
Ref.~\cite{marleyPRC}.

In the comparison shown in Fig.~\ref{fig:Ev_resolution}, solid lines represent
the \marleyNew\ results, while dashed lines of the same color are used to plot
the \marleyOld\ results. Below a neutrino energy of about \SI{12}{\MeV}, the
behavior of the two \marley\ versions is similar apart from a lower threshold
in \marleyNew\ for nonzero $f_\mathrm{RMS}$ values for the
$E_{e+\gamma}^\mathrm{reco}$ (dark orange), $E_\mathrm{vis}^\mathrm{reco}$
(dark blue), and $E_\mathrm{tag}^\mathrm{p}$ (cyan) energy estimators. This
change reflects a slightly lower threshold for neutron emission under the
\marleyNew\ treatment of continuum nuclear transitions. At moderate neutrino
energies between about \SI{12}{\MeV} and \SI{35}{\MeV}, all of the estimators
show some improvement in energy resolution. This is especially noticeable for
$E_e^\mathrm{reco}$ (green), for which the solid curve falls well below the
dashed curve in this region. The improvement can be attributed to the
suppression of the allowed strength in the \marleyNew\ model arising from
corrections for nonzero $\kappa$, which are most important for excitation
energies above the threshold for emitting nuclear fragments.

At the highest neutrino energies shown, the decreasing trend of all \marleyOld\
curves is replaced by an increase in \marleyNew. This behavior is a consequence
of the inclusion of forbidden continuum transitions in the new \marley\ model.
In the AA treatment from \marleyOld, once the neutrino energy is high enough to
allow all major GT transitions to be populated, the typical excitation energy
imparted to the nucleus saturates. Further increases to $E_\nu$ simply increase
the energy of the outgoing electron, leading to an underestimation of the role
of nuclear de-excitation products in the overall energy budget. This problem is
avoided in \marleyNew\ via inclusion of continuum forbidden transitions. Under
the new \marley\ model, larger neutrino energies will allow progressively
higher-order multipoles to contribute to the cross section. Since the giant
resonances corresponding to these forbidden multipole transitions are expected
to have a centroid excitation energy of roughly $41\,J\,A^{-1/3}$
\si{\MeV}~\cite{Kolbe:2003ys} (where $J$ is the multipole order and $A = 40$ is
the nucleon number), the typical energy transfer to the nucleus will continue
to increase with $E_\nu$, and channels involving a higher multiplicity of
emitted nuclear fragments will become correspondingly more important.

\section{Summary and conclusions}
\label{sec:summary}

As the anticipated operation of the DUNE far detector draws nearer, increasing
attention is being focused on its potential as a low-energy neutrino
observatory. As the only neutrino event generator that provides a realistic
model of the $\isotope[40]{Ar}(\nu_e, e^{-})\isotope[40]{K}^*$ reaction at the
sub-\SI{100}{\MeV} energies of interest, the \marley\ software package has been
widely used as a tool to study LArTPC sensitivity to astrophysical neutrinos.
To overcome key limitations in the physics model used in the current
version~\oldVersion\ of \marley, we have implemented major enhancements to its
theoretical treatment of the primary neutrino interaction. These enhancements
have focused on removal of the AA, which was previously the basis for \marley\
predictions of the inclusive cross section. In the unbound continuum of nuclear
excitation energies, the previous \marley\ model has been completely replaced
with an HF-CRPA calculation, including both allowed and forbidden multipoles up
to order $J = 5$. For allowed transitions to low-lying discrete nuclear levels,
corrections have been added to the original model that account for $1/m_N$
terms in the nuclear current, nucleon form factors, and the momentum transfer
dependence of the nuclear matrix elements. In addition to these cross-section
adjustments, the \marley\ nuclear de-excitation model has also been upgraded to
use the recent KDUQFederal evaluation of the nuclear optical
potential~\cite{Pruitt2023}, and the continuum threshold has been adjusted to
be more consistent with the one used for the primary interaction. All of these
improvements will be shared with the community in an upcoming open-source
release of version~\version\ of the \marley\ code.

Many comparisons are reported in this article between the updated \marley\
predictions and those from version~\oldVersion. In most cases, the observables
in these comparisons have been chosen to match those reported in a previous
\marley\ publication~\cite{marleyPRC}. Throughout this work, the allowed
transitions are seen to remain dominant at few tens-of-\si{\MeV} energies after
our model refinements, but their overall strength is substantially reduced from
the AA estimate. Due to the updates to the continuum threshold in the
de-excitation model as well as the inclusion of forbidden transitions, the
cross section for the exclusive $1n1p$ final state is notably enhanced compared
to prior calculations. Because the reduction in the allowed portion of the
cross section is especially strong at backward angles (which correspond to
relatively high values of the momentum transfer $\kappa$), the electron angular
distribution generated by supernova neutrino interactions is predicted to be
more concentrated in the forward direction than expected from the AA. This
feature may allow a future DUNE analysis to leverage the CC
$\nu_e$-\isotope[40]{Ar} reaction for supernova pointing. The much larger
statistics in this channel may lead to improved pointing resolution over the
existing strategy, which relies on isolating neutrino-electron elastic
scattering events~\cite{DUNE:2024ptd}.

Although this work has greatly increased the theoretical sophistication of the
\marley\ neutrino cross-section model, the code still relies on key
approximations that may be examined in future research. Although their
contribution is expected to be small, forbidden transitions to discrete nuclear
levels are still entirely neglected in \marley~\version. These notably include
the transition to the $4^-$ ground state of \isotope[40]{K}. As a result, the
$\nu_e$-\isotope[40]{Ar} CC reaction threshold in \marley\ is known to be
overestimated by roughly \SI{2}{\MeV}. For the discrete transitions that are
presently included in the updated \marley\ model, the use of the nuclear form
factor $F(\kappa)$ discussed in Sec.~\ref{sec:nuclear_form_factor} assumes that
the radial dependence of the nuclear matrix elements is the same as the
\isotope[40]{Ar} nuclear density and does not vary between $\isotope[40]{K}$
energy levels. While the use of $F(\kappa)$ is far superior to the lack of any
similar correction in the AA, it remains a rough approximation that could be
tested and refined through dedicated calculations of the matrix elements in the
discrete level region. As noted previously in Ref.~\cite{marleyPRC}, the
validity of the \marley\ nuclear de-excitation model rests on the assumption
that the primary neutrino interaction immediately forms a compound nuclear
state. While we have made no attempt herein to move beyond this simplifying
approximation, further study of the possible role of direct fragment knockout
and pre-equilibrium nuclear de-excitations would also be well-motivated.

The theoretical formalism developed in this work is suitable for predictions of
CC cross sections for both neutrinos and antineutrinos, but results are
presented only for the former. While the HF-CRPA approach has already been used
to study antineutrino-argon CC scattering~\cite{VanDessel:2019atx,
VanDessel:2019obk}, completion of a corresponding \marley-based treatment
requires additional inputs. For the $\isotope[40]{Ar}(\bar{\nu}_e,
e^{+})\isotope[40]{Cl}^*$ reaction, the Fermi matrix element from
Eq.~\ref{eq:BF} vanishes due to isospin selection rules; there is no isobaric
analog in \isotope[40]{Cl} of the \isotope[40]{Ar} ground state. Therefore, one
cannot use the method from Sec.~\ref{sec:crpa_energy_balance} to determine the
value of the energy shift parameter $\DeltaCRPA$ for the
$\bar{\nu}_e$-\isotope[40]{Ar} CC interaction. A greater difficulty is posed by
the lack of measured GT matrix elements needed to describe antineutrino-induced
transitions to discrete \isotope[40]{Cl} energy levels, including the two known
$1^+$ levels at excitation energies of \SI{0.8895}{\MeV} and
\SI{2.3072}{\MeV}~\cite{Chen:2017ngq}. New experimental data to probe the
relevant GT strength in \isotope[40]{Cl}, as might be obtained via
charge-exchange reactions using techniques similar to those discussed in
Refs.~\cite{Bhattacharya2009, Karakoc2014}, would be invaluable to maximize the
quality of a future \marley\ implementation of the
$\bar{\nu}_e$-\isotope[40]{Ar} CC process.

Beyond serving as essential input to a complete simulation of low-energy
$\bar{\nu}_e$ CC primary interactions for DUNE and similar experiments, the GT
strength measurements proposed here would also enable a particularly sensitive
test of the full \marley\ interaction model. Although the CC cross section has
not yet been measured for an \isotope[40]{Ar} target using either neutrinos or
antineutrinos at energies relevant for \marley, muon capture data are already
available for the total rate, prompt de-excitation $\gamma$-ray
intensities~\cite{mucapGammas}, and the yields of several final-state
nuclides~\cite{mucapar, mucapar2}. Since the muon capture process typically
involves a tens-of-\si{\MeV} energy transfer to the nucleus and shares the same
nuclear matrix elements with antineutrino CC scattering via crossing symmetry,
future \marley\ comparisons to these data would provide powerful model
constraints.

Benchmarking the \marley~\version\ model against muon capture data would be
complementary to similar studies that could be carried out using present and
anticipated low-energy neutrino cross-section measurements. While a full
\marleyNew\ calculation for iodine and lead is reserved for future work, we
note that the recent COHERENT $\mu$DAR neutrino cross-section measurements for
these nuclear targets~\cite{ninspb, coherent:2023ffx} both indicate a
substantial overestimation in the \marleyOld\ model for final states containing
at least one neutron. Updated predictions for these specific nuclei will need
to be completed before firm conclusions can be made about the origin of these
discrepancies. However, the large suppression of the high-lying allowed
strength seen here in the updated \marley\ model for \isotope[40]{Ar} is
expected to be a generic feature for other targets. It is therefore plausible
that this effect may play a significant role in explaining the tension between
the earlier \marley\ model and these COHERENT results.

With the use of the CEvNS nuclear form factor to evaluate $F(\kappa)$ in the
\marley~\version\ treatment of discrete transitions, a more realistic
simulation for that neutral-current (NC) process becomes straightforward to
implement. By combining discrete neutral GT strengths based on photonuclear
interaction data~\cite{Tornow:2025rwj} (which have been used previously to
estimate neutrino-argon NC cross sections~\cite{newmark:2023vup,
Meighen-Berger:2024xbx}) with HF-CRPA NC responses~\cite{VanDessel:2019atx,
VanDessel:2019obk} we also plan to develop a comprehensive \marley\ model of
inelastic NC reactions on \isotope[40]{Ar} in the near future.

\section{Acknowledgments}

The authors thank the Mainz Institute for Theoretical Physics for hosting the
June 2023 topical workshop ``Neutrino Scattering at Low and Intermediate
Energies''~\cite{mitpWorkshop}. The workshop facilitated early discussions
between us that helped to develop this work.

S.G. received support under award number DE-SCL0000022 from the Early Career
Research Program, U.S. Department of Energy, Office of Science, Office of High
Energy Physics. Participation of L.H.A. in this work was supported by the
Science Undergraduate Laboratory Internships (SULI) program, U.S. Department of
Energy, Office of Science, Office of Workforce Development for Teachers and
Scientists (WDTS). A.N. is supported by the Neutrino Theory Network under Award Number DEAC02-07CH1135. N.J. received support from the Ghent University Special Research Fund (BOF) and the Fund for Scientific Research Flanders (FWO-Flanders).

This manuscript has been authored by Fermi Forward Discovery Group, LLC under
Contract No. 89243024CSC000002 with the U.S. Department of Energy, Office of
Science, Office of High Energy Physics.

More than one billion \marley\ events were produced for this study using services provided by the OSG Consortium~\cite{osg07, osg09, ospool, osdf}, which is supported by the National Science Foundation awards \#2030508 and \#2323298.

\appendix

\section{Basic operators}
\label{sec:basic_operators}

When working in the Walecka formalism~\cite{Walecka1975} to first order in
$1/m_N$, the nuclear matrix elements needed to evaluate neutrino-nucleus cross
sections can be expressed in terms of eight basic spherical tensor operators,
as shown in Eqs.~\ref{eq:first_Walecka_op}--\ref{eq:last_Walecka_op}. In the
notation used in this article, all of the basic operators $T \in \{ M, \Delta,
\Delta^{\prime}, \Delta^{\prime\prime}, \Sigma, \Sigma^{\prime},
\Sigma^{\prime\prime}, \Omega \}$ involve a sum over single-nucleon
contributions via
\begin{equation}
T_J^M = \sum_{n = 1}^{A} T_J^M(n) \, t_{\mp}(n) \,,
\end{equation}
where the symbol $(n)$ immediately following an operator is used to indicate
that the operator acts only on the $n$-th nucleon.

The single-nucleon version of the first basic operator is given by
\begin{equation}
M_J^M(\nucIndex) \equiv j_J( \kappa \, r_\nucIndex )
  \, Y_{JM}(\hat{\mathbf{r}}_\nucIndex) \,,
\end{equation}
where $j_J$ is the $J$-th spherical Bessel function of the first kind and
$Y_{JM}$ is a spherical harmonic function. Using the definitions in the
appendix of Ref.~\cite{Serot:1978vj} as well as Eqs. (45), (47), and (48) from
Ref.~\cite{OConnell:1972edu} (which follow from the properties of the $\nabla$
operator), one may write the remaining seven single-nucleon basic operators in
terms of $M_J^M(\nucIndex)$ as shown below.
\onecolumngrid

\begin{equation}
\Delta_J^M(\nucIndex) = \frac{ i }{ \kappa }
  \, \big[ M_J \otimes \mathbf{p} \big]_{JM}(\nucIndex)
\end{equation}

\begin{equation}
{\Delta^{\prime}}_J^M(\nucIndex) =
  \frac{ i }{ \kappa } \left( \frac{ J + 1 }{ 2J + 1 } \right)^{1/2}
    \! \big[ M_{J-1} \otimes \mathbf{p} \big]_{JM}(\nucIndex)
  - \frac{ i }{ \kappa } \left( \frac{ J }{ 2J + 1 } \right)^{1/2}
    \! \big[ M_{J+1} \otimes \mathbf{p} \big]_{JM}(\nucIndex)
\end{equation}

\begin{equation}
{\Delta^{\prime\prime}}_J^M(\nucIndex) =
  \frac{ i }{ \kappa } \left( \frac{ J + 1 }{ 2J + 1 } \right)^{1/2}
    \! \big[ M_{J+1} \otimes \mathbf{p} \big]_{JM}(\nucIndex)
  + \frac{ i }{ \kappa } \left( \frac{ J }{ 2J + 1 } \right)^{1/2}
    \! \big[ M_{J-1} \otimes \mathbf{p} \big]_{JM}(\nucIndex)
\end{equation}

\begin{equation}
\Sigma_J^M(\nucIndex) = \big[ M_J \otimes \boldsymbol{\sigma} \big]_{JM}(\nucIndex)
\end{equation}

\begin{equation}
{\Sigma^{\prime}}_J^M(\nucIndex) =
  \left( \frac{ J + 1 }{ 2J + 1 } \right)^{1/2}
    \! \big[ M_{J-1} \otimes \boldsymbol{\sigma} \big]_{JM}(\nucIndex)
  - \left( \frac{ J }{ 2J + 1 } \right)^{1/2}
    \! \big[ M_{J+1} \otimes \boldsymbol{\sigma} \big]_{JM}(\nucIndex)
\end{equation}

\begin{equation}
{\Sigma^{\prime\prime}}_J^M(\nucIndex) =
  \left( \frac{ J + 1 }{ 2J + 1 } \right)^{1/2}
    \! \big[ M_{J+1} \otimes \boldsymbol{\sigma} \big]_{JM}(\nucIndex)
  + \left( \frac{ J }{ 2J + 1 } \right)^{1/2}
    \! \big[ M_{J-1} \otimes \boldsymbol{\sigma} \big]_{JM}(\nucIndex)
\end{equation}

\begin{equation}
\Omega_J^M(\nucIndex) = \frac{ i }{ \kappa } \, M_J^M(\nucIndex)
  \, \boldsymbol{\sigma}(\nucIndex) \cdot \mathbf{p}(\nucIndex)
\end{equation}
\twocolumngrid

Here $\mathbf{p} = -i\nabla$ is the momentum operator, and the product of
spherical tensors $A$ and $B$ with ranks $K$ and $L$, respectively, is given by
\begin{equation}
\big[ A \otimes B \big]_{JM}
 = \sum_{u} \sum_{v} \big< K \, u \, L \, v \, | \, J \, M \big>
  \, A_u \, B_v
\end{equation}
where $A_u$ ($B_v$) is the $u$-th ($v$-th) component of $A$ ($B$). For example,
\begin{align}
\nonumber
\big[ M_J \otimes \boldsymbol{\sigma} &\big]_{JM}(\nucIndex) \\
  &= \sum_{u} \sum_{v} \big< J \, u \, 1 \, v \, | \, J \, M \big>
    \, M_J^u(\nucIndex) \, \sigma_v(\nucIndex) \,,
\end{align}
since the Pauli vector $\boldsymbol{\sigma}$ is a rank-1 spherical tensor.

The basic operators are written in
Eqs.~\ref{eq:first_Walecka_op}--\ref{eq:last_Walecka_op} without their $^M$
superscripts because they appear in matrix elements from
Eqs.~\ref{eq:first_Walecka_ME}--\ref{eq:last_Walecka_ME} that have been reduced
using the Wigner-Eckhart theorem.

\bibliography{main.bib}

\end{document}